\documentclass[aps,epsfigure,twocolumn,superscriptaddress,nofootinbib]{revtex4-1}
\usepackage[brazil]{babel}
\usepackage[utf8]{inputenc}
\usepackage{makeidx} 
\usepackage{graphicx} 
\usepackage{dcolumn} 
\usepackage{array} 
\usepackage{amssymb} 
\usepackage{amsmath}
\usepackage{textcomp}
\usepackage{multirow}
\usepackage{subfigure}
\usepackage{eucal}
\usepackage{mathrsfs}
\usepackage[all]{xy}
\usepackage{epstopdf}
\usepackage{color}
\usepackage{float} 
\usepackage{amsmath} 
\usepackage{amsfonts} 
\usepackage{indentfirst} 
\usepackage[breakable]{tcolorbox}
    \makeatletter
    \newcommand{\boxspacing}{\kern\kvtcb@left@rule\kern\kvtcb@boxsep}
    \makeatother

 \definecolor{incolor}{HTML}{303F9F}
    \definecolor{outcolor}{HTML}{D84315}
    \definecolor{cellborder}{HTML}{CFCFCF}
    \definecolor{cellbackground}{HTML}{F7F7F7}
  
 \usepackage{upquote} 
    \usepackage{eurosym} 
    \usepackage[mathletters]{ucs} 
    \usepackage{fancyvrb} 
    \usepackage{grffile} 
    \makeatletter 
    \def\Gread@@xetex#1{%
      \IfFileExists{"\Gin@base".bb}%
      {\Gread@eps{\Gin@base.bb}}%
      {\Gread@@xetex@aux#1}%
    }
    \makeatother
 
    \DefineVerbatimEnvironment{Highlighting}{Verbatim}{commandchars=\\\{\}}

    \DefineVerbatimEnvironment{Highlighting}{Verbatim}{commandchars=\\\{\}}

    \let\Oldtex\TeX
    \let\Oldlatex\LaTeX
    \renewcommand{\TeX}{\textrm{\Oldtex}}
    \renewcommand{\LaTeX}{\textrm{\Oldlatex}}
    \title{artigo_qiskit}

\makeatletter
\def\PY@reset{\let\PY@it=\relax \let\PY@bf=\relax%
    \let\PY@ul=\relax \let\PY@tc=\relax%
    \let\PY@bc=\relax \let\PY@ff=\relax}
\def\PY@tok#1{\csname PY@tok@#1\endcsname}
\def\PY@toks#1+{\ifx\relax#1\empty\else%
    \PY@tok{#1}\expandafter\PY@toks\fi}
\def\PY@do#1{\PY@bc{\PY@tc{\PY@ul{%
    \PY@it{\PY@bf{\PY@ff{#1}}}}}}}
\def\PY#1#2{\PY@reset\PY@toks#1+\relax+\PY@do{#2}}

\expandafter\def\csname PY@tok@w\endcsname{\def\PY@tc##1{\textcolor[rgb]{0.73,0.73,0.73}{##1}}}
\expandafter\def\csname PY@tok@c\endcsname{\let\PY@it=\textit\def\PY@tc##1{\textcolor[rgb]{0.25,0.50,0.50}{##1}}}
\expandafter\def\csname PY@tok@cp\endcsname{\def\PY@tc##1{\textcolor[rgb]{0.74,0.48,0.00}{##1}}}
\expandafter\def\csname PY@tok@k\endcsname{\let\PY@bf=\textbf\def\PY@tc##1{\textcolor[rgb]{0.00,0.50,0.00}{##1}}}
\expandafter\def\csname PY@tok@kp\endcsname{\def\PY@tc##1{\textcolor[rgb]{0.00,0.50,0.00}{##1}}}
\expandafter\def\csname PY@tok@kt\endcsname{\def\PY@tc##1{\textcolor[rgb]{0.69,0.00,0.25}{##1}}}
\expandafter\def\csname PY@tok@o\endcsname{\def\PY@tc##1{\textcolor[rgb]{0.40,0.40,0.40}{##1}}}
\expandafter\def\csname PY@tok@ow\endcsname{\let\PY@bf=\textbf\def\PY@tc##1{\textcolor[rgb]{0.67,0.13,1.00}{##1}}}
\expandafter\def\csname PY@tok@nb\endcsname{\def\PY@tc##1{\textcolor[rgb]{0.00,0.50,0.00}{##1}}}
\expandafter\def\csname PY@tok@nf\endcsname{\def\PY@tc##1{\textcolor[rgb]{0.00,0.00,1.00}{##1}}}
\expandafter\def\csname PY@tok@nc\endcsname{\let\PY@bf=\textbf\def\PY@tc##1{\textcolor[rgb]{0.00,0.00,1.00}{##1}}}
\expandafter\def\csname PY@tok@nn\endcsname{\let\PY@bf=\textbf\def\PY@tc##1{\textcolor[rgb]{0.00,0.00,1.00}{##1}}}
\expandafter\def\csname PY@tok@ne\endcsname{\let\PY@bf=\textbf\def\PY@tc##1{\textcolor[rgb]{0.82,0.25,0.23}{##1}}}
\expandafter\def\csname PY@tok@nv\endcsname{\def\PY@tc##1{\textcolor[rgb]{0.10,0.09,0.49}{##1}}}
\expandafter\def\csname PY@tok@no\endcsname{\def\PY@tc##1{\textcolor[rgb]{0.53,0.00,0.00}{##1}}}
\expandafter\def\csname PY@tok@nl\endcsname{\def\PY@tc##1{\textcolor[rgb]{0.63,0.63,0.00}{##1}}}
\expandafter\def\csname PY@tok@ni\endcsname{\let\PY@bf=\textbf\def\PY@tc##1{\textcolor[rgb]{0.60,0.60,0.60}{##1}}}
\expandafter\def\csname PY@tok@na\endcsname{\def\PY@tc##1{\textcolor[rgb]{0.49,0.56,0.16}{##1}}}
\expandafter\def\csname PY@tok@nt\endcsname{\let\PY@bf=\textbf\def\PY@tc##1{\textcolor[rgb]{0.00,0.50,0.00}{##1}}}
\expandafter\def\csname PY@tok@nd\endcsname{\def\PY@tc##1{\textcolor[rgb]{0.67,0.13,1.00}{##1}}}
\expandafter\def\csname PY@tok@s\endcsname{\def\PY@tc##1{\textcolor[rgb]{0.73,0.13,0.13}{##1}}}
\expandafter\def\csname PY@tok@sd\endcsname{\let\PY@it=\textit\def\PY@tc##1{\textcolor[rgb]{0.73,0.13,0.13}{##1}}}
\expandafter\def\csname PY@tok@si\endcsname{\let\PY@bf=\textbf\def\PY@tc##1{\textcolor[rgb]{0.73,0.40,0.53}{##1}}}
\expandafter\def\csname PY@tok@se\endcsname{\let\PY@bf=\textbf\def\PY@tc##1{\textcolor[rgb]{0.73,0.40,0.13}{##1}}}
\expandafter\def\csname PY@tok@sr\endcsname{\def\PY@tc##1{\textcolor[rgb]{0.73,0.40,0.53}{##1}}}
\expandafter\def\csname PY@tok@ss\endcsname{\def\PY@tc##1{\textcolor[rgb]{0.10,0.09,0.49}{##1}}}
\expandafter\def\csname PY@tok@sx\endcsname{\def\PY@tc##1{\textcolor[rgb]{0.00,0.50,0.00}{##1}}}
\expandafter\def\csname PY@tok@m\endcsname{\def\PY@tc##1{\textcolor[rgb]{0.40,0.40,0.40}{##1}}}
\expandafter\def\csname PY@tok@gh\endcsname{\let\PY@bf=\textbf\def\PY@tc##1{\textcolor[rgb]{0.00,0.00,0.50}{##1}}}
\expandafter\def\csname PY@tok@gu\endcsname{\let\PY@bf=\textbf\def\PY@tc##1{\textcolor[rgb]{0.50,0.00,0.50}{##1}}}
\expandafter\def\csname PY@tok@gd\endcsname{\def\PY@tc##1{\textcolor[rgb]{0.63,0.00,0.00}{##1}}}
\expandafter\def\csname PY@tok@gi\endcsname{\def\PY@tc##1{\textcolor[rgb]{0.00,0.63,0.00}{##1}}}
\expandafter\def\csname PY@tok@gr\endcsname{\def\PY@tc##1{\textcolor[rgb]{1.00,0.00,0.00}{##1}}}
\expandafter\def\csname PY@tok@ge\endcsname{\let\PY@it=\textit}
\expandafter\def\csname PY@tok@gs\endcsname{\let\PY@bf=\textbf}
\expandafter\def\csname PY@tok@gp\endcsname{\let\PY@bf=\textbf\def\PY@tc##1{\textcolor[rgb]{0.00,0.00,0.50}{##1}}}
\expandafter\def\csname PY@tok@go\endcsname{\def\PY@tc##1{\textcolor[rgb]{0.53,0.53,0.53}{##1}}}
\expandafter\def\csname PY@tok@gt\endcsname{\def\PY@tc##1{\textcolor[rgb]{0.00,0.27,0.87}{##1}}}
\expandafter\def\csname PY@tok@err\endcsname{\def\PY@bc##1{\setlength{\fboxsep}{0pt}\fcolorbox[rgb]{1.00,0.00,0.00}{1,1,1}{\strut ##1}}}
\expandafter\def\csname PY@tok@kc\endcsname{\let\PY@bf=\textbf\def\PY@tc##1{\textcolor[rgb]{0.00,0.50,0.00}{##1}}}
\expandafter\def\csname PY@tok@kd\endcsname{\let\PY@bf=\textbf\def\PY@tc##1{\textcolor[rgb]{0.00,0.50,0.00}{##1}}}
\expandafter\def\csname PY@tok@kn\endcsname{\let\PY@bf=\textbf\def\PY@tc##1{\textcolor[rgb]{0.00,0.50,0.00}{##1}}}
\expandafter\def\csname PY@tok@kr\endcsname{\let\PY@bf=\textbf\def\PY@tc##1{\textcolor[rgb]{0.00,0.50,0.00}{##1}}}
\expandafter\def\csname PY@tok@bp\endcsname{\def\PY@tc##1{\textcolor[rgb]{0.00,0.50,0.00}{##1}}}
\expandafter\def\csname PY@tok@fm\endcsname{\def\PY@tc##1{\textcolor[rgb]{0.00,0.00,1.00}{##1}}}
\expandafter\def\csname PY@tok@vc\endcsname{\def\PY@tc##1{\textcolor[rgb]{0.10,0.09,0.49}{##1}}}
\expandafter\def\csname PY@tok@vg\endcsname{\def\PY@tc##1{\textcolor[rgb]{0.10,0.09,0.49}{##1}}}
\expandafter\def\csname PY@tok@vi\endcsname{\def\PY@tc##1{\textcolor[rgb]{0.10,0.09,0.49}{##1}}}
\expandafter\def\csname PY@tok@vm\endcsname{\def\PY@tc##1{\textcolor[rgb]{0.10,0.09,0.49}{##1}}}
\expandafter\def\csname PY@tok@sa\endcsname{\def\PY@tc##1{\textcolor[rgb]{0.73,0.13,0.13}{##1}}}
\expandafter\def\csname PY@tok@sb\endcsname{\def\PY@tc##1{\textcolor[rgb]{0.73,0.13,0.13}{##1}}}
\expandafter\def\csname PY@tok@sc\endcsname{\def\PY@tc##1{\textcolor[rgb]{0.73,0.13,0.13}{##1}}}
\expandafter\def\csname PY@tok@dl\endcsname{\def\PY@tc##1{\textcolor[rgb]{0.73,0.13,0.13}{##1}}}
\expandafter\def\csname PY@tok@s2\endcsname{\def\PY@tc##1{\textcolor[rgb]{0.73,0.13,0.13}{##1}}}
\expandafter\def\csname PY@tok@sh\endcsname{\def\PY@tc##1{\textcolor[rgb]{0.73,0.13,0.13}{##1}}}
\expandafter\def\csname PY@tok@s1\endcsname{\def\PY@tc##1{\textcolor[rgb]{0.73,0.13,0.13}{##1}}}
\expandafter\def\csname PY@tok@mb\endcsname{\def\PY@tc##1{\textcolor[rgb]{0.40,0.40,0.40}{##1}}}
\expandafter\def\csname PY@tok@mf\endcsname{\def\PY@tc##1{\textcolor[rgb]{0.40,0.40,0.40}{##1}}}
\expandafter\def\csname PY@tok@mh\endcsname{\def\PY@tc##1{\textcolor[rgb]{0.40,0.40,0.40}{##1}}}
\expandafter\def\csname PY@tok@mi\endcsname{\def\PY@tc##1{\textcolor[rgb]{0.40,0.40,0.40}{##1}}}
\expandafter\def\csname PY@tok@il\endcsname{\def\PY@tc##1{\textcolor[rgb]{0.40,0.40,0.40}{##1}}}
\expandafter\def\csname PY@tok@mo\endcsname{\def\PY@tc##1{\textcolor[rgb]{0.40,0.40,0.40}{##1}}}
\expandafter\def\csname PY@tok@ch\endcsname{\let\PY@it=\textit\def\PY@tc##1{\textcolor[rgb]{0.25,0.50,0.50}{##1}}}
\expandafter\def\csname PY@tok@cm\endcsname{\let\PY@it=\textit\def\PY@tc##1{\textcolor[rgb]{0.25,0.50,0.50}{##1}}}
\expandafter\def\csname PY@tok@cpf\endcsname{\let\PY@it=\textit\def\PY@tc##1{\textcolor[rgb]{0.25,0.50,0.50}{##1}}}
\expandafter\def\csname PY@tok@c1\endcsname{\let\PY@it=\textit\def\PY@tc##1{\textcolor[rgb]{0.25,0.50,0.50}{##1}}}
\expandafter\def\csname PY@tok@cs\endcsname{\let\PY@it=\textit\def\PY@tc##1{\textcolor[rgb]{0.25,0.50,0.50}{##1}}}

\def\PYZus{\char`\_}

\def\PYZgt{\char`\>}
\def\PYZsh{\char`\#}
\def\PYZpc{\char`\%}

\def\PYZhy{\char`\-}

\def\PYZdq{\char`\"}

\makeatother

    \makeatletter
        \newbox\Wrappedcontinuationbox 
        \newbox\Wrappedvisiblespacebox 
        \newcommand*\Wrappedvisiblespace {\textcolor{red}{\textvisiblespace}} 
        \newcommand*\Wrappedcontinuationsymbol {\textcolor{red}{\llap{\tiny$\m@th\hookrightarrow$}}} 
        \newcommand*\Wrappedcontinuationindent {3ex } 
        \newcommand*\Wrappedafterbreak {\kern\Wrappedcontinuationindent\copy\Wrappedcontinuationbox} 
        \newcommand*\Wrappedbreaksatspecials {%
            \def\PYGZus{\discretionary{\char`\_}{\Wrappedafterbreak}{\char`\_}}%
            \def\PYGZob{\discretionary{}{\Wrappedafterbreak\char`\{}{\char`\{}}%
            \def\PYGZcb{\discretionary{\char`\}}{\Wrappedafterbreak}{\char`\}}}%
            \def\PYGZca{\discretionary{\char`\^}{\Wrappedafterbreak}{\char`\^}}%
            \def\PYGZam{\discretionary{\char`\&}{\Wrappedafterbreak}{\char`\&}}%
            \def\PYGZlt{\discretionary{}{\Wrappedafterbreak\char`\<}{\char`\<}}%
            \def\PYGZgt{\discretionary{\char`\>}{\Wrappedafterbreak}{\char`\>}}%
            \def\PYGZsh{\discretionary{}{\Wrappedafterbreak\char`\#}{\char`\#}}%
            \def\PYGZpc{\discretionary{}{\Wrappedafterbreak\char`\%}{\char`\%}}%
            \def\PYGZdl{\discretionary{}{\Wrappedafterbreak\char`\$}{\char`\$}}%
            \def\PYGZhy{\discretionary{\char`\-}{\Wrappedafterbreak}{\char`\-}}%
            \def\PYGZsq{\discretionary{}{\Wrappedafterbreak\textquotesingle}{\textquotesingle}}%
            \def\PYGZdq{\discretionary{}{\Wrappedafterbreak\char`\"}{\char`\"}}%
            \def\PYGZti{\discretionary{\char`\~}{\Wrappedafterbreak}{\char`\~}}%
        } 
        \newcommand*\Wrappedbreaksatpunct {%
            \lccode`\~`\.\lowercase{\def~}{\discretionary{\hbox{\char`\.}}{\Wrappedafterbreak}{\hbox{\char`\.}}}%
            \lccode`\~`\,\lowercase{\def~}{\discretionary{\hbox{\char`\,}}{\Wrappedafterbreak}{\hbox{\char`\,}}}%
            \lccode`\~`\;\lowercase{\def~}{\discretionary{\hbox{\char`\;}}{\Wrappedafterbreak}{\hbox{\char`\;}}}%
            \lccode`\~`\:\lowercase{\def~}{\discretionary{\hbox{\char`\:}}{\Wrappedafterbreak}{\hbox{\char`\:}}}%
            \lccode`\~`\?\lowercase{\def~}{\discretionary{\hbox{\char`\?}}{\Wrappedafterbreak}{\hbox{\char`\?}}}%
            \lccode`\~`\!\lowercase{\def~}{\discretionary{\hbox{\char`\!}}{\Wrappedafterbreak}{\hbox{\char`\!}}}%
            \lccode`\~`\/\lowercase{\def~}{\discretionary{\hbox{\char`\/}}{\Wrappedafterbreak}{\hbox{\char`\/}}}%
            \catcode`\.\active
            \catcode`\,\active 
            \catcode`\;\active
            \catcode`\:\active
            \catcode`\?\active
            \catcode`\!\active
            \catcode`\/\active 
            \lccode`\~`\~ 	
        }
    \makeatother

    \let\OriginalVerbatim=\Verbatim
    \makeatletter
    \renewcommand{\Verbatim}[1][1]{%
        \sbox\Wrappedcontinuationbox {\Wrappedcontinuationsymbol}%
        \sbox\Wrappedvisiblespacebox {\FV@SetupFont\Wrappedvisiblespace}%
        \def\FancyVerbFormatLine ##1{\hsize\linewidth
            \vtop{\raggedright\hyphenpenalty\z@\exhyphenpenalty\z@
                \doublehyphendemerits\z@\finalhyphendemerits\z@
                \strut ##1\strut}%
        }%
        \def\FV@Space {%
            \nobreak\hskip\z@ plus\fontdimen3\font minus\fontdimen4\font
            \discretionary{\copy\Wrappedvisiblespacebox}{\Wrappedafterbreak}
            {\kern\fontdimen2\font}%
        }%
        
        \Wrappedbreaksatspecials
        \OriginalVerbatim[#1,codes*=\Wrappedbreaksatpunct]%
    }
    \makeatother

\begin{document}

\title{Computação Quântica: uma abordagem para a graduação usando o Qiskit \\
  \small Quantum Computing: an undergraduate approach using Qiskit}

\author{Gleydson Fernandes de Jesus}
\affiliation{Grupo de Informa\c{c}\~{a}o Qu\^{a}ntica, Centro de Ci\^{e}ncias Exatas e das Tecnologias, Universidade Federal do Oeste da Bahia - Campus Reitor Edgard Santos. Rua Bertioga, 892, Morada Nobre I, 47810-059 Barreiras, Bahia, Brasil.}
\author{Maria Heloísa Fraga da Silva}
\affiliation{Grupo de Informa\c{c}\~{a}o Qu\^{a}ntica, Centro de Ci\^{e}ncias Exatas e das Tecnologias, Universidade Federal do Oeste da Bahia - Campus Reitor Edgard Santos. Rua Bertioga, 892, Morada Nobre I, 47810-059 Barreiras, Bahia, Brasil.}

\author{Teonas Gonçalves Dourado Netto}
\affiliation{Grupo de Informa\c{c}\~{a}o Qu\^{a}ntica, Centro de Ci\^{e}ncias Exatas e das Tecnologias, Universidade Federal do Oeste da Bahia - Campus Reitor Edgard Santos. Rua Bertioga, 892, Morada Nobre I, 47810-059 Barreiras, Bahia, Brasil.}

\author{Lucas Queiroz Galvão}
\affiliation{Grupo de Informa\c{c}\~{a}o Qu\^{a}ntica, Centro de Ci\^{e}ncias Exatas e das Tecnologias, Universidade Federal do Oeste da Bahia - Campus Reitor Edgard Santos. Rua Bertioga, 892, Morada Nobre I, 47810-059 Barreiras, Bahia, Brasil.}

\author{Frankle Gabriel de Oliveira Souza}
\affiliation{Grupo de Informa\c{c}\~{a}o Qu\^{a}ntica, Centro de Ci\^{e}ncias Exatas e das Tecnologias, Universidade Federal do Oeste da Bahia - Campus Reitor Edgard Santos. Rua Bertioga, 892, Morada Nobre I, 47810-059 Barreiras, Bahia, Brasil.}

\author{Clebson Cruz}
\email{clebson.cruz@ufob.edu.br}
\affiliation{Grupo de Informa\c{c}\~{a}o Qu\^{a}ntica, Centro de Ci\^{e}ncias Exatas e das Tecnologias, Universidade Federal do Oeste da Bahia - Campus Reitor Edgard Santos. Rua Bertioga, 892, Morada Nobre I, 47810-059 Barreiras, Bahia, Brasil.}

\date{\today}

\begin{abstract}
Neste artigo, apresentamos a ferramenta Quantum Information Software Developer Kit - \textit{Qiskit}, para o ensino de computação quântica para estudantes de graduação, com conhecimento básico dos postulados da mecânica quântica. Nos concentramos na apresentação da construção dos programas em qualquer laptop ou desktop comum e a sua execução em processadores quânticos reais através do acesso remoto aos \textit{hardwares} disponibilizados na plataforma \textit{IBM Quantum Experience}. Os códigos são disponibilizados ao longo do texto para que os leitores, mesmo com pouca experiência em computação científica, possam reproduzí-los e adotar os métodos discutidos neste artigo para abordar seus próprios projetos de computação quântica. Os resultados apresentados estão de acordo com as previsões teóricas e mostram a eficácia do pacote \textit{Qiskit} como uma ferramenta de trabalho em sala de aula, robusta para a introdução de conceitos aplicados de computação e informação quântica. 

\noindent
\textbf{Palavras-Chave:}Python, IBM, Qiskit, Quantum Experience.

\vspace{10pt}

In this paper, we present the Quantum Information Software Developer Kit - \textit{Qiskit}, for teaching quantum computing to undergraduate students, with basic knowledge of quantum mechanics postulates. We focus on presenting the construction of the programs on any common laptop or desktop computer and their execution on real quantum processors through the remote access to the \textit{quantum hardware} available on the \textit{IBM Quantum Experience} platform. The codes are made available throughout the text so that readers, even with little experience in scientific computing, can reproduce them and adopt the methods discussed in this paper to address their own quantum computing projects. The results presented are in agreement with theoretical predictions and show the effectiveness of the \textit{Qiskit} package as a robust classroom working tool for the introduction of applied concepts of quantum computing and quantum information theory.

\end{abstract}

\maketitle

\section{Introdução}

Com o advento do \textit{IBM Quantum Experience} (IBM QE) \cite{ibm,alves2020simulating,santos2017computador,rabelo2018abordagem}, houve uma facilitação ao acesso a plataformas de computação quântica \cite{nielsen2002quantum,scarani1998quantum,vedral1998basics,steane1998quantum,bennett2000quantum,oliveira2020fisica} por qualquer pessoa com acesso à internet através de um computador doméstico \cite{santos2017computador,rabelo2018abordagem}.
Entretanto, a maioria dos estudantes dos cursos de ciências exatas e tecnológicas não são apresentados aos conceitos fundamentais da computação quântica até a pós-graduação. Muitos desses estudantes são fascinados com conceitos de Computação Quântica \cite{nielsen2002quantum,scarani1998quantum,vedral1998basics,steane1998quantum,bennett2000quantum,oliveira2020fisica}, uma vez que, antes mesmo de ingressarem em um curso de graduação, já estão familiarizados com o fato de que computadores quânticos superam o poder de processamento dos computadores comerciais disponíveis atualmente. Nos últimos, anos os avanços mostrados pela computação quântica têm mostrado o seu o potencial de revolução tecnológica \cite{terhal2018quantum,harrow2017quantum}. Nesse cenário, a computação científica dos próximos anos será liderada por aqueles que têm o conhecimento acerca da utilização de dispositivos quânticos. Portanto, se torna crucial facilitar o acesso à educação quântica, garantindo que os estudantes, independentemente de planejarem trabalhar em uma área relacionada à teoria da informação quântica, aprendam conceitos básicos de computação quântica.

Nesse contexto, a apresentação da área de Computação e Informação Quântica para alunos de graduação tem atraído a atenção da comunidade científica nos últimos anos \cite{scarani1998quantum,benenti2008quantum,jose2013introduccao,candela2015undergraduate,fedortchenko2016quantum,santos2017computador,rabelo2018abordagem,alves2020simulating,castillo2020classical,perry2019quantum}. Além de contextualizar o processo de ensino e aprendizagem no cotidiano, alguns estudos apontam que o uso de tecnologias como recurso auxiliar de aprendizagem constitui uma realidade para a maior parte dos estudantes, sendo um caminho profíquo para a consolidação  do que se compreende como democratização e universalização do conhecimento \cite{teixeira,faria}, de modo que trabalhos recentes têm apostado na plataforma IBM QE como aliada das práticas pedagógicas, propondo, inclusive, abordagens didáticas para o ensino de computação quântica no nível de graduação \cite{santos2017computador,rabelo2018abordagem,alves2020simulating} e até mesmo no ensino médio \cite{perry2019quantum,tappert2019experience}. 

Em 2017 a IBM (International Business Machines) disponibilizou o seu kit de desenvolvimento de software para informação quântica (Quantum Information Software Developer Kit), ou simplesmente \textit{Qiskit} \cite{qiskit-textbook,qiskit-documentation,github-qiskit,qiskit,larose2019overview}, permitindo o desenvolvimento de softwares para seu serviço de computação quântica em nuvem \cite{ibm}. As contribuições podem ser feitas por apoiadores externos, através da plataforma GitHub \cite{github-qiskit}, onde são disponibilizados uma série de exemplos de algoritmos quânticos da comunidade \cite{qiskit-tutorials} e trazem um conjunto de exercícios que auxiliam no aprendizado de computação quântica \cite{qiskit-textbook}.

Neste trabalho, analisamos o pacote \textit{Qiskit}, usando a linguagem Python 3 \cite{python,oliphant2007python,van2011python}, como um recurso educacional para aulas de computação quântica para a graduação em Física e áreas afins, além do desenvolvimento de potenciais projetos de pesquisa e iniciação científica sênior e júnior. Mostramos como essa pode ser uma ferramenta poderosa para o ensino de computação quântica, com foco na implementação de circuitos quânticos simples e algoritmos quânticos bem conhecidos. Apresentamos as principais condições para a construção dos programas e a sua execução em processadores quanticos reais, ou até mesmo em computadores domésticos. Os códigos são disponibilizados nos Boxes ao longo do texto, de modo que os leitores possam adotar os métodos discutidos neste artigo para abordar seus próprios projetos de computação quântica. 

Vale destacar que, este artigo traz um resumo das notas de aula da disciplina CET0448 - Tópicos Especiais III: Computação Quântica Aplicada, ministrada para estudantes do primeiro ao último semestre dos cursos de Licenciatura e Bacharelado em Física da Universidade Federal do Oeste da Bahia.  O pacote \textit{Qiskit} foi utilizado como uma ferramenta de trabalho para a apresentação de conceitos básicos de computação e informação quântica para uma ampla gama de estudantes, com um conhecimento básico de mecânica quântica e nenhuma experiência em programação científica. 

Esse trabalho está estruturado seguindo um roteiro básico para a introdução de conceitos fundamentais para a computação quântica como qubits, portas quânticas, emaranhamento e algoritmos quânticos, seguindo a estrutura apresentada na disciplina. Primeiramente, na Seção 2 fazemos uma apresentação das ferramentas computacionais necessárias para abordar nossos projetos de computação quântica em computadores domésticos; na seção 3 comentamos brevemente os principais conceitos básicos de computação e informação quântica como bits quânticos, portas quânticas básicas, medidas e emaranhamento quântico; na Seção 4 apresentamos as aplicações, fornecendo um conjunto de problemas abordados pelos estudantes da disciplina (autores deste trabalho), executados em processadores quânticos reais. Nessa seção, apresentaremos a construção de portas lógicas clássicas a partir de portas quânticas, o famoso algoritmo de teleporte quântico \cite{santos2017computador,rabelo2018abordagem,nielsen2002quantum,oliveira2020fisica} e o algoritmo de busca de Grover \cite{nielsen2002quantum,castillo2020classical}. Estas aplicações podem ser usadas como exemplos de implementação de algoritmos quânticos, apresentando o \textit{Qiskit} como uma ferramenta de trabalho útil para o ensino de computação quântica. Finalizamos o trabalho com as conclusões na seção 5.


\section{Ferramentas Computacionais}
\label{compframe}

A linguagem Python \cite{python,oliphant2007python,van2011python} foi projetada para ser de fácil leitura, com rápido desenvolvimento de código e de fácil compreensão, pois tem pouco foco na sintaxe e um foco maior nos conceitos básicos de lógica de programação \cite{oliphant2007python}. No entanto, apesar da flexibilidade, o Python é considerada lenta em comparação com outras linguagens, mas isso é compensado por sua biblioteca robusta e fácil de manipular, adequada para cálculos científicos \cite{python,oliphant2007python,van2011python,kadiyala2017applications,harris2020array}. Nesse quesito, a utilização do Python para o \textit{Qiskit} permite que os conteúdos apresentados neste artigo possam ser reproduzidos pela maioria dos leitores, mesmo aqueles que têm pouca ou nenhuma experiência com essa linguagem de programação.

\subsection{Jupyter Notebook e Anaconda(Python)}
Recomendamos que os leitores usem o \textit{software} livre Jupyter Notebook \cite{kluyver2016jupyter,jupyter,glick2018using,cardoso2018using,zuniga2020digital,perkel2018jupyter,hussain2018introducing} em seus projetos Python de computação quântica, especialmente aqueles que não têm experiência em computação científica ou estão começando a aprender a linguagem Python. Recentemente, diversos trabalhos têm apontado a eficácia do Jupyter Notebook para o aprendizado de computação de alta performace \cite{glick2018using,cardoso2018using,zuniga2020digital,perkel2018jupyter,hussain2018introducing}. O Jupyter Notebook facilita a interação entre o usuário e o computador, permitindo a inclusão de textos na formatação \LaTeX\ e a apresentação de resultados gráficos durante a execução dos programas, permitindo ao usuário acompanhar em tempo real cada etapa do código, auxiliando na compreensão dos códigos e seus resultados, sendo uma ferramenta robusta para o ensino de computação quântica. Além disso, uma das vantagens no uso do Jupyter é que o \textit{IBM QE} \cite{ibm} usa um ambiente Jupyter Notebook, que permite programar com Python na nuvem usando o pacote Qiskit em um computador quântico real a partir de um computador doméstico,  e até mesmo emular um processador quântico a partir da unidade de processamento local do usuário mesmo sem acesso a internet. 

O Jupyter Notebook, assim como o Python 3, podem ser facilmente encontrados para download gratuito na internet. Ambos fazem parte de uma das plataformas de ciência de dados mais populares da atualidade, o Anaconda \cite{anaconda,kadiyala2017applications,perkel2018jupyter,hussain2018introducing}.

O Anaconda\footnote{A última versão do Anaconda (4.8.3) pode ser baixada gratuitamente no site da plataforma \cite{anaconda}, baseado no sistema operacional do computador do usuário. Nós recomendamos utilizar a instalação padrão. Depois de instalado, o usuário pode abrir o Anaconda Navigator no seu computador e atestar que a instalação foi concluída com êxito.} é uma ferramenta computacional que vem completamente pronta para uso, sendo um ambiente de desenvolvimento para várias linguagens populares, como Python, C, Java, R, Julia, entre outras \cite{anaconda}. O Anaconda vem  com todas as bibliotecas necessárias para modelar sistemas físicos como numpy, scipy e matplotlib, entre outros (150) pacotes pré-instalados e mais de 250 pacotes de código aberto que podem ser adicionados \cite{kadiyala2017applications}. Dentre esses pacotes disponíveis para o repositório do Anaconda encontramos o \textit{Qiskit} \cite{qiskit}, elemento fundamental para esse trabalho. A seguir, mostramos uma breve introdução ao pacote \textit{Qiskit}.

\subsection{Quantum Information Software Developer Kit -\textit{Qiskit}}

O Quantum Information Software Developer Kit -\textit{Qiskit}\footnote{A forma recomendada de instalar o \textit{Qiskit} é utilizando o gerenciador de pacotes do Python, (\texttt{pip}), pré-instalado nas últimas versões do Python e Anaconda, utilizando o comando no terminal \texttt{>\ pip\ install\ qiskit}. Para uma instalação detalhada, recomendamos acessar a seção de instalação na página do github dos projetos \cite{github-qiskit}.} \cite{qiskit-textbook,qiskit-documentation,github-qiskit,qiskit,larose2019overview} é uma estrutura computacional de código aberto desenvolvida para funcionar em diferentes linguagens de programação como Python \cite{github-qiskit}, Swift \cite{qiskit-swift} e JavaScript \cite{qiskit-js}, fornecendo as ferramentas necessárias para a criação de algoritmos quânticos, seguindo um modelo de circuito para computação quântica universal \cite{nielsen2002quantum}, e a sua execução em dispositivos quânticos reais usando o acesso remoto aos \textit{hardwares} disponibilizados através do IBM QE. Além disso, o Qiskit permite emular um computador quântico em  processador clássico local, como um laptop ou um desktop comum, permitindo a testagem de algoritmos quânticos simples em qualquer computador doméstico, sem a necessidade de acesso à internet ou criação de uma conta no IBM QE.

O IBM QE oferece a estudantes, pesquisadores e entusiastas da computação quântica acesso rápido e prático por meio de uma interface amigável, permitindo que os usuários executem seus projetos e experimentos \cite{ibm,santos2017computador,rabelo2018abordagem}. Por outro lado, o \textit{Qiskit} é uma ferramenta profisional para o desenvolvimento de programação quântica de alto nível \cite{qiskit-textbook,qiskit-documentation,github-qiskit,qiskit,larose2019overview}, sendo tanto uma plataforma de desenvolvimento de \textit{softwares} quânticos como uma linguagem de programação quântica \cite{larose2019overview}. Para isso, o \textit{Qiskit} conta com cinco elementos essenciais:
\begin{itemize}
 \item[]\textbf{\textit{Terra:}} contém os elementos fundamentais que são usados para escrever os circuitos dos algoritmos quânticos;
 \item[]\textbf{\textit{Aer:}} contém os recursos para as simulações quânticas por meio de computação de alto desempenho;
 \item[]\textbf{\textit{AQUA:}} algoritmos para aplicativos de computação quântica, ou \textit{AQUA}, fornece as bibliotecas para aplicativos específicos de algoritmos, como Química, Finanças e Machine Learning. 
 \item[]\textbf{\textit{Ignis:}} contém ferramentas específicas para algoritmos de correção de erros, ruídos quânticos e verificação de hardware quântico.
 \item[]\textbf{\textit{IBM Q Provider:}} não é necessariamente um elemento fundamental, mas fornece as ferramentas para acessar \textit{IBM Q Experience}, a fim de executar programas de usuários em um processador quântico real.
\end{itemize}

Neste artigo usamos o \textit{Qiskit} na linguagem Python 3  para construir os circuitos quânticos e para as simulações dos algoritmos em computadores quânticos reais, usando apenas os elementos \textit{Terra}, \textit{Aer} e \textit{IBM Q Provider}.

\subsection{Importando os Pacotes}

Uma vez instalados o Anaconda (Python) e o \textit{Qiskit} em seus computadores, os usuários estão prontos para aprender como escrever códigos para simular seus próprios algoritmos quânticos, construindo circuitos e executando-os em seus próprios computadores domésticos. Para iniciar o programa, é necessário adicionar estes recursos no ambiente Python no Jupyter Notebook, importando os seguintes módulos:
\begin{itemize}
 \item[] \texttt{qiskit}: para projetar os circuitos quânticos e executar algoritmos quânticos \cite{qiskit-textbook,qiskit-documentation,github-qiskit,qiskit};
 \item[] \texttt{numpy}: para construir um ambiente matemático com arrays e matrizes multidimensionais, usando sua grande coleção de funções matemáticas \cite{harris2020array};
 \item[] \texttt{matplotlib}: para a criação de gráficos e visualizações de dados em geral \cite{hunter2007matplotlib};
 \item[] \texttt{qiskit.tools.monitor}: para utilizarmos a função \texttt{job{\_}monitor} para monitorar em tempo real a execução dos nossos algoritmos \cite{qiskit-textbook,qiskit-documentation,github-qiskit,qiskit};
 \item[] \texttt{qiskit.visualization}: para utilizar as funções \texttt{plot{\_}histogram} para visualizar os resultados através das distribuições de probabilidade e \texttt{plot{\_}bloch{\_}}\texttt{multivector} para visualizar os estados na representação da esfera de bloch \cite{nielsen2002quantum}.
\end{itemize}

Esses módulos básicos podem ser importados logo na primeira célula do notebook do Jupyter e executado com o comando \texttt{shift+enter} no teclado\footnote{As células do Jupyter Notebook são sempre executadas através do comando \texttt{shift+enter} no teclado.}, sempre que um novo notebook for criado. Para isso, usamos os seguintes comandos:
\begin{tcolorbox}[breakable, size=fbox, boxrule=1pt, pad at break*=1mm,colback=cellbackground, colframe=cellborder,coltitle=black,title=Box 1: Importando os Pacotes]
\begin{Verbatim}[commandchars=\\\{\}]
\PY{k+kn}{from} \PY{n+nn}{qiskit} \PY{k+kn}{import} \PY{o}{*}
\PY{k+kn}{import} \PY{n+nn}{numpy} \PY{k}{as} \PY{n+nn}{np}
\PY{k+kn}{import} \PY{n+nn}{matplotlib}\PY{n+nn}{.}\PY{n+nn}{pyplot} \PY{k}{as} \PY{n+nn}{plt}
\PY{k+kn}{from} \PY{n+nn}{qiskit}\PY{n+nn}{.}\PY{n+nn}{tools}\PY{n+nn}{.}\PY{n+nn}{monitor} \PY{k+kn}{import} \PY{n}{job\PYZus{}monitor}
\PY{k+kn}{from} \PY{n+nn}{qiskit}\PY{n+nn}{.}\PY{n+nn}{visualization} \PY{k+kn}{import} \PY{n}{plot\PYZus{}histogram}
\PY{k+kn}{from} \PY{n+nn}{qiskit}\PY{n+nn}{.}\PY{n+nn}{visualization} \PY{k+kn}{import} \PY{n}{plot\PYZus{}bloch\PYZus{}multivector}
\PY{o}{\PYZpc{}}\PY{k}{matplotlib} inline
\end{Verbatim}
\end{tcolorbox}

Vale destacar que, o comando \texttt{\%matplotlib inline} serve para definir o processo interno do \texttt{matplotlib}, permitindo que as saídas dos comandos de plotagem seja exibida de forma embutida na interface frontal, como o Jupyter Notebook,  abaixo da célula em que o código é escrito \cite{hunter2007matplotlib}.

Uma vez que os pacotes estão importados, temos todas as condições de começar a programar algoritmos quânticos em nosso computador pessoal e executá-los de forma remota nos computadores quânticos disponibilizados pela IBM \cite{ibm,alves2020simulating,santos2017computador,rabelo2018abordagem}.

\section{Fundamentos:}
\label{theoretical}

Nesta seção, fornecemos uma breve introdução aos conceitos fundamentais de informação quântica e computação quântica, usando os ambientes computacionais descritos na última seção. Descrevemos os conceitos de qubits, emaranhamento quântico, portas lógicas quânticas, circuitos e algoritmos. Esses tópicos foram amplamente estudados e discutidos na literatura nas últimas décadas \cite{nielsen2002quantum,bennett2000quantum,steane1998quantum,vedral1998basics,vedral2006introduction,gyongyosi2019survey}. Para os leitores que têm somente um conhecimento básico em mecânica quântica, recomendamos a leitura complementar das referências \cite{santos2017computador,rabelo2018abordagem,alves2020simulating,castillo2020classical,perry2019quantum,oliveira2020fisica}. Para leitores com conhecimento avançado em mecânica quântica, recomendamos as referências~\cite{nielsen2002quantum,vedral2006introduction,bennett2000quantum} para uma descrição mais detalhada dos tópicos abordados nesta seção.

\subsection{Bits Quânticos (Qubits)}

BInary DigiT, ou \textit{bit} é a menor unidade de informação em uma teoria da informação clássica, e a teoria da computação clássica é fundamentada neste conceito \cite{candela2015undergraduate,nielsen2002quantum}. O \textit{bit} clássico é um estado lógico que assume um dos dois valores possíveis $\lbrace 0,1\rbrace$. Outras representações úteis, são $\lbrace \texttt{sim},\texttt{não}\rbrace$,  $\lbrace \texttt{verdadeiro},\texttt{falso}\rbrace$ ou $\lbrace \texttt{ligado},\texttt{desligado}\rbrace$. Em computadores clássicos, essas duas possibilidades podem ser implementadas usando componentes eletrônicos clássicos de dois estados, como dois níveis de tensão ou corrente distintos e estáveis em um circuito, duas posições de interruptores elétricos, dois níveis de intensidade de luz ou polarização e dois estados elétricos diferentes de um circuito flip-flop \cite{pedroni2008digital}, por exemplo. Assim, os computadores são projetados com instruções para manipular e armazenar múltiplos \textit{bits}, chamados bytes (conjunto de 8 bits).

Da mesma forma, a teoria da informação quântica e a computação quântica são construídas através de uma unidade de informação fundamental, análoga ao \textit{bit} (clássico): os bits quânticos, ou simplesmente qubits \cite{nielsen2002quantum}. No entanto, enquanto os bits clássicos podem assumir uma das duas possibilidades acima mencionadas, os qubits podem ser representados como uma combinação linear da base ortonormal de um sistema quântico de dois níveis, convencionalmente representada como  $\lbrace \vert{0}\rangle,\vert{1}\rangle\rbrace$, chamada de \textit{base computacional} \cite{nielsen2002quantum,santos2017computador}, onde em uma representação matricial:
\begin{eqnarray}
\vert{0}\rangle = \begin{bmatrix} 1 \\ 0 \end{bmatrix} \label{eq:01}\\
\vert{1}\rangle = \begin{bmatrix} 0 \\ 1 \end{bmatrix} \label{eq:02}
\end{eqnarray}

Portanto, a principal vantagem dos qubits sobre os bits está no princípio de sobreposição \cite{nielsen2002quantum,vedral2006introduction,oliveira2020fisica,griffiths2018introduction} o que possibilita combinações lineares entre os vetores que compõem a base computacional. Desta forma, a representação mais geral para um qubit é um vetor $\vert{\psi}\rangle$ escrito como:
\begin{eqnarray}
\vert{\psi}\rangle = \alpha \vert{0}\rangle + \beta \vert{1}\rangle ~, 
\label{eq:03}
\end{eqnarray}
onde $\alpha$ e $\beta$ são amplitudes complexas que obedecem à condição de normalização $\vert \alpha\vert ^{2} + \vert \beta\vert ^{2}=1$, com $\vert \alpha\vert ^{2}$ corresponde à probabilidade de obter o estado  $\vert{0}\rangle$ e $\vert \beta\vert ^{2}$ a probabilidade de obter o estado $\vert{1}\rangle$, através de uma medida no estado $\vert{\psi}\rangle$.

Após a importação dos pacotes apresentadas no Box 1, temos todas as condições de criar o conjunto de regras ou operações que, aplicadas nos qubits, permitem solucionar algum problema preestabelecido, ou seja, os algoritmos quânticos. 

O primeiro passo é definir as bases do circuito que implementará o algoritmo desejado, começando pelo conjunto de qubits que será utilizado no problema. Para isso, definimos uma variável\footnote{O nome das váriáveis é de livre escolha do usuário.} \texttt{q} usando a função \texttt{QuantumRegister} da seguinte forma:

\begin{tcolorbox}[breakable, size=fbox, boxrule=1pt, pad at break*=1mm,colback=cellbackground, colframe=cellborder,coltitle=black,title=Box 2: Registrando os qubits]
\begin{Verbatim}[commandchars=\\\{\}]
\PY{n}{q} \PY{o}{=} \PY{n}{QuantumRegister}\PY{p}{(}\PY{n}{N}\PY{p}{,} \PY{l+s+s1}'\PY{l+s+s1}{q}\PY{l+s+s1}'\PY{p}{)}
\end{Verbatim}
\end{tcolorbox}

No Box 2, \texttt{N} é um número inteiro e representa o número de qubits que será usado no circuito. Por definição, os qubits são sempre registrados no estado $\vert{0}\rangle^{\otimes N}$, ou seja, cada um dos \texttt{N} qubits no estado $\vert{0}\rangle$. 

Outro elemento importante na construção do circuito quântico é a definição do conjunto de bits clássicos onde registramos a informação oriunda das medidas realizadas nos qbits, após a execução de algum algoritmo, por exemplo. Para isso, de maneira análoga aos qubits, definimos uma variável \texttt{b} usando a função \texttt{ClassicalRegister}:
\begin{tcolorbox}[breakable, size=fbox, boxrule=1pt, pad at break*=1mm,colback=cellbackground, colframe=cellborder,coltitle=black,title=Box 3: Registrando os bits clássicos]
\begin{Verbatim}[commandchars=\\\{\}]
\PY{n}{b} \PY{o}{=} \PY{n}{ClassicalRegister}\PY{p}{(}\PY{n}{N}\PY{p}{,} \PY{l+s+s1}'\PY{l+s+s1}{b}\PY{l+s+s1}'\PY{p}{)}
\end{Verbatim}
\end{tcolorbox}

Finalmente, podemos então declarar a variável \texttt{circuito} para construir o nosso circuito usando o conjunto de bits clássicos e quânticos definidos anteriormente através da função \texttt{QuantumCircuit}:
\begin{tcolorbox}[breakable, size=fbox, boxrule=1pt, pad at break*=1mm,colback=cellbackground, colframe=cellborder,coltitle=black,title=Box 4: Criando o circuito]
\begin{Verbatim}[commandchars=\\\{\}]
\PY{n}{circuito} \PY{o}{=} \PY{n}{QuantumCircuit}\PY{p}{(}\PY{n}{qubits}\PY{p}{,} \PY{n}{bits}\PY{p}{)}
\end{Verbatim}
\end{tcolorbox}

Nesse ponto temos a base para o nosso circuito e temos todas as condições de definir os três componentes pincipais de todo algoritmo quântico:
\begin{itemize}
 \item[]\textbf{\textit{Inicialização:}} Primeiro, precisamos iniciar nosso processo de computação em um estado bem definido.
 \item[]\textbf{\textit{Portas Quânticas:}} Em seguida, aplicamos a sequência de operações (portas) quânticas que  permitem solucionar o problema preestabelecido;
 \item[]\textbf{\textit{Medidas:}} Finalizamos, medindo os estados de cada qubit, registramos as medidas nos bits clássicos, e usando um computador clássico, interpretamos as medições através das distribuições de probabilidade correspondente a cada resultado das medidas.
\end{itemize}

A seguir apresentamos cada etapa da construção de um algoritmo quântico.

\subsection{Inicialização}

Usando o \textit{Qiskit} podemos definir os coeficientes $\alpha$ e $\beta$ e inicializar cada qubit do circuito no estado descrito na equação~(\ref{eq:03}). Para isso usamos os seguintes comandos:
\begin{tcolorbox}[breakable, size=fbox, boxrule=1pt, pad at break*=1mm,colback=cellbackground, colframe=cellborder,coltitle=black,title=Box 5: Inicializando um qubit em um determinado estado $\vert{\psi}\rangle$]
\begin{Verbatim}[commandchars=\\\{\}]
\PY{n}{psi} \PY{o}{=} \PY{p}{[}\PY{n}{alpha}\PY{p}{,}\PY{n}{beta}\PY{p}{]}
\PY{n}{circuito}\PY{o}{.}\PY{n}{initialize}\PY{p}{(}\PY{n}{psi}\PY{p}{,}\PY{n}{q}\PY{p}{[}\PY{n}{i}\PY{p}{]}\PY{p}{)}
\end{Verbatim}
\end{tcolorbox}
Onde a variável \texttt{psi} é uma matriz que representa o estado descrito na equação~(\ref{eq:03}), com as variáveis \texttt{alpha} e \texttt{beta} correspondendo aos coeficientes $\alpha$ e $\beta$, respectivamente, e \texttt{q[i]} o qubit \texttt{q} índice \texttt{i} que será inicializado no estado $\vert{\psi}\rangle$.

\subsubsection{Esfera de Bloch}

Nesse ponto, vale destacar uma representação útil para o estado de um qubit, que pode ser obtida através do mapeamento das componentes  $\alpha$ e $\beta$ como funções de ângulos $\theta$ e $\phi$. Dessa maneira, devido ao fato de  $\alpha$ e $\beta$ obedecerem à condição de normalização $\vert \alpha\vert ^{2} + \vert \beta\vert ^{2}=1$, equação~(\ref{eq:03}) pode ser reescrita como
\begin{equation}
\vert{\psi}\rangle = \cos\left(\frac{\theta}{2}\right) \vert{0}\rangle + e^{i\phi}\sin\left(\frac{\theta}{2}\right) \vert{1}\rangle~, 
\label{eq:04}
\end{equation}

Assim, o par $\lbrace\theta,\phi\rbrace$ define um ponto em uma esfera de raio unitário, conhecida na literatura como \textit{Esfera de Bloch} \cite{nielsen2002quantum}, que nos dá uma representação geométrica para o espaço de Hilbert de um qubit. 

Nessa representação, o estado de um qubit corresponde a um ponto na superfície da esfera de Bloch e estados ortogonais são diametralmente opostos\footnote{Isso explica o fato de usarmos $\frac{\theta}{2}$ na equação~(\ref{eq:04})}. Através da importação do pacote \texttt{qiskit.visualization}, previamente instalado junto ao \textit{Qiskit}, podemos usar a função \texttt{plot{\_}bloch{\_}}\texttt{multivector} para obtermos a visualização do qubit de interesse na esfera de Bloch. 

Assim, escolhendo o par $\lbrace\theta,\phi\rbrace$ na equação~(\ref{eq:04}), podemos obter a representação geométrica do qubit descrito por $\vert{\psi}\rangle$. Vamos analizar a inicialização dos qubits através de alguns exemplos. Primeiramente, importamos os pacotes necessários conforme descrito no Box 1; em seguida registramos um qubit (\texttt{N}=1) conforme descrito no Box 2; criamos um circuito conforme o Box 4, sem a necessidade de um bit clássico auxiliar, pois não serão feitas medidas nesse qubit. Finalmente, podemos inicializar nosso qubit a partir dos ângulos  $\theta$ e $\phi$. Usando o pacote \texttt{numpy} (chamado por \texttt{np}), definimos os coeficientes $\alpha$ e $\beta$ a partir dos ângulos $\theta$ e $\phi$ e, conforme apresentado no Box 5, inicializamos o nosso estado. Todo esse processo é apresentado no Box 6, a seguir:
\begin{tcolorbox}[breakable, size=fbox, boxrule=1pt, pad at break*=1mm,colback=cellbackground, colframe=cellborder,coltitle=black,title=Box 6: Inicializar o qubit a partir dos ângulos  $\theta$ e $\phi$]
\begin{Verbatim}[commandchars=\\\{\}]
\PY{n}{theta} \PY{o}{=} \PY{p}{(}\PY{n+nb}{float}\PY{p}{(}\PY{n+nb}{input}\PY{p}{(}\PY{l+s+s2}{\PYZdq{}}\PY{l+s+s2}{Insira o ângulo theta(°): }\PY{l+s+s2}{\PYZdq{}}\PY{p}{)}\PY{p}{)}\PY{p}{)}\PY{o}{*}\PY{n}{np}\PY{o}{.}\PY{n}{pi}\PY{o}{/}\PY{p}{(}\PY{l+m+mi}{180}\PY{p}{)}
\PY{n}{phi}   \PY{o}{=} \PY{p}{(}\PY{n+nb}{float}\PY{p}{(}\PY{n+nb}{input}\PY{p}{(}\PY{l+s+s2}{\PYZdq{}}\PY{l+s+s2}{Insira o ângulo phi(°): }\PY{l+s+s2}{\PYZdq{}}\PY{p}{)}\PY{p}{)}\PY{p}{)}\PY{o}{*}\PY{n}{np}\PY{o}{.}\PY{n}{pi}\PY{o}{/}\PY{p}{(}\PY{l+m+mi}{180}\PY{p}{)}
\PY{n}{alpha} \PY{o}{=} \PY{n}{np}\PY{o}{.}\PY{n}{cos}\PY{p}{(}\PY{n}{theta}\PY{o}{/}\PY{l+m+mi}{2}\PY{p}{)}
\PY{n}{beta} \PY{o}{=} \PY{p}{(}\PY{n}{np}\PY{o}{.}\PY{n}{exp}\PY{p}{(}\PY{l+m+mi}{1}\PY{n}{j}\PY{o}{*}\PY{n}{phi}\PY{p}{)}\PY{p}{)}\PY{o}{*}\PY{n}{np}\PY{o}{.}\PY{n}{sin}\PY{p}{(}\PY{n}{theta}\PY{o}{/}\PY{l+m+mi}{2}\PY{p}{)}
\PY{n}{estado\PYZus{}inicial} \PY{o}{=} \PY{p}{[}\PY{n}{alpha}\PY{p}{,}\PY{n}{beta}\PY{p}{]}
\PY{n}{circuito}\PY{o}{.}\PY{n}{initialize}\PY{p}{(}\PY{n}{estado\PYZus{}inicial}\PY{p}{,}\PY{n}{qubit}\PY{p}{[}\PY{l+m+mi}{0}\PY{p}{]}\PY{p}{)}
\end{Verbatim}
\end{tcolorbox}

Finalmente, podemos usar o elemento \texttt{Aer} do \textit{Qiskit} para simular o estado inicializado em nosso computador local, obter o vetor de estado e plotá-lo na representação da esfera de Bloch, usando o pacote \texttt{plot{\_}} \texttt{bloch{\_}}\texttt{multivector}
 
\begin{tcolorbox}[breakable, size=fbox, boxrule=1pt, pad at break*=1mm,colback=cellbackground, colframe=cellborder,coltitle=black,title=Box 7: Plotando o qubit na Esfera de Bloch]
\begin{Verbatim}[commandchars=\\\{\}]
\PY{n}{processo} \PY{o}{=} \PY{n}{Aer}\PY{o}{.}\PY{n}{get\PYZus{}backend}\PY{p}{(}\PY{l+s+s1}'\PY{l+s+s1}{statevector\PYZus{}simulator}\PY{l+s+s1}'\PY{p}{)}
\PY{n}{vector\PYZus{}de\PYZus{}estado}  \PY{o}{=} \PY{n}{execute}\PY{p}{(}\PY{n}{circuito}\PY{p}{,} \PY{n}{backend}\PY{o}{=}\PY{n}{processo}\PY{p}{)}\PY{o}{.}\PY{n}{result}\PY{p}{(}\PY{p}{)}\PY{o}{.}\PY{n}{get\PYZus{}statevector}\PY{p}{(}\PY{p}{)}
\PY{n}{plot\PYZus{}bloch\PYZus{}multivector}\PY{p}{(}\PY{n}{vector\PYZus{}de\PYZus{}estado}\PY{p}{)}
\end{Verbatim}
\end{tcolorbox}

\begin{figure}[h]
    \centering
    \subfigure[]{\includegraphics[scale=0.45]{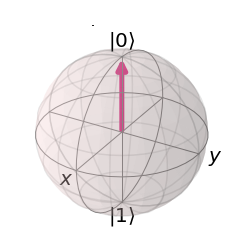}}\qquad
    \subfigure[]{\includegraphics[scale=0.45]{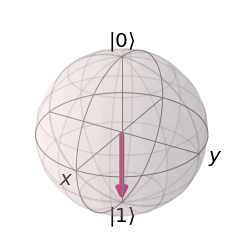}} \\
    \subfigure[]{\includegraphics[scale=0.45]{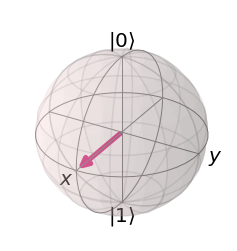}}\qquad
    \subfigure[]{\includegraphics[scale=0.45]{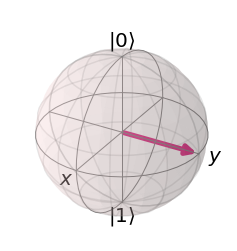}}
    \caption{Representação da esfera de Bloch de um qubit. Escolhendo os ângulos $\theta$ e $\phi$ na equação~(\ref{eq:04}) obtemos a representação da esfera Blcoh para  os estados (a) $\vert{\psi}\rangle=\vert{0}\rangle$ ($\theta = 0^{o}$); (b) $\vert{\psi}\rangle=\vert{1}\rangle$ ($\theta = 180^{o}$); (c) $\vert{\psi}\rangle =\vert{+}\rangle =\frac{1}{\sqrt{2}} \left( \vert{0}\rangle + \vert{1}\rangle\right)$ $\lbrace \theta=90^{o}$, $\phi=0^{o} \rbrace$; e $\vert{\psi}\rangle =\vert{+i}\rangle =\frac{1}{\sqrt{2}} \left( \vert{0}\rangle + i\vert{1}\rangle\right)$ $\lbrace \theta=90^{o}$, $\phi=90^{o} \rbrace$.}
    \label{fig:01}
\end{figure}

A fig.~\ref{fig:01} mostra a representação da esfera Bloch para qubits inicializados em ângulos específicos. 

Um outro caminho para a inicialização é aplicação de operações que transformam o sistema de qubits inicialmente registrado no estado $\vert{0}\rangle ^{\otimes N}$. Essas operações são conhecidas como portas quânticas.

\subsection{Portas Quânticas}

Uma vez definido o elemento básico de informação quântica (os qubits), temos todas as condições de introduzir os conjuntos de operações que atuam sobre eles. Em computação clássica, essas operações são implementadas pelo que conhecemos como portas lógicas \cite{oliveira2020fisica}. As Portas Lógicas Clássicas seguem uma Álgebra Booleana \cite{whitesitt2012boolean} e são implementadas a partir de circuitos eletrônicos \cite{oliveira2020fisica}, geralmente usando diodos ou transistores que atuam como interruptores eletrônicos, permitindo a implementação de alguma operação lógica através de uma determinada função booleana \cite{whitesitt2012boolean}. Assim, essas portas são aplicadas em circuitos lógicos para a implementação de processos computacionais, levando a solução de problemas através de algoritmos.

Na Computação Quântica, analogamente à computação clássica, o conjunto de operações que atuam sobre os qubits são conhecidos como Portas Lógicas Quânticas, ou simplesmente Portas Quânticas. Ao contrário das portas lógicas clássicas, as portas quânticas são sempre reversíveis \cite{oliveira2020fisica}\footnote{Em computação clássica a única porta reversível é a porta \texttt{NOT}  \cite{oliveira2020fisica}.}. Devido à grande quantidade de portas quânticas e às suas semelhanças de implementação no \textit{Qiskit}, apresentamos a seguir as principais Portas Quânticas que utilizaremos ao longo desse trabalho, em sua forma matricial.

\subsubsection{Portas de 1 qubit} 

Vamos começar com as portas quânticas de 1 qubit, a partir do que conhecemos como portas quânticas elementares, ou portas de Pauli \cite{oliveira2020fisica}, que correspondem às matrizes de Pauli \cite{oliveira2020fisica,griffiths2018introduction}:
\begin{eqnarray}
X &=&  \begin{bmatrix} 0 & 1 \\ 1 & 0 \end{bmatrix}~, \label{eq:05} \\
Y &=&  \begin{bmatrix} 0 & -i \\ i & 0 \end{bmatrix}~, \label{eq:06} \\
Z &=&  \begin{bmatrix} 1 & 0 \\ 0 & -1 \end{bmatrix}~. \label{eq:07} 
\end{eqnarray}

Consideremos o estado descrito na equação~(\ref{eq:03}). A atuação dessas portas nesse estado é:
\begin{eqnarray}
X\vert{\psi}\rangle &=&   \alpha \vert{1}\rangle +\beta \vert{0}\rangle ~, \label{eq:08} \\
Y\vert{\psi}\rangle &=&  i\alpha \vert{1}\rangle -i\beta \vert{0}\rangle ~, \label{eq:09} \\
Z\vert{\psi}\rangle &=&   \alpha \vert{0}\rangle -\beta \vert{1}\rangle ~. \label{eq:010} 
\end{eqnarray}

Assim, pode-se perceber que as portas de Pauli correspondem a uma rotação na esfera de Bloch de $\pi$~rad no eixo correspondente à direção representada pela porta.

Uma outra porta muito importante, e que compõe o conjunto de portas quânticas universais \cite{oliveira2020fisica} - através da qual qualquer transformação unitária pode ser implementada em um estado quântico genérico - é a porta de fase ou porta $S$, onde:

\begin{eqnarray}
S &=&  \begin{bmatrix} 1 & 0 \\ 0 & i \end{bmatrix}~.
\end{eqnarray}
Consideremos novamente o estado descrito na equação~(\ref{eq:03}). A atuação dessas portas nesse estado é:
\begin{eqnarray}
S \vert{\psi}\rangle &=&   \alpha \vert{0}\rangle +i\beta \vert{1}\rangle ~. \label{eq:010} 
\end{eqnarray}

Na literatura \cite{nielsen2002quantum,scarani1998quantum,vedral1998basics,steane1998quantum,bennett2000quantum,oliveira2020fisica}, a porta de fase é comumente conhecida como porta $\sqrt{Z}$, isso porque a aplicação da porta S duas vezes consecutivas equivale a aplicação da porta $Z$.

\textbf{Porta \texttt{NOT}  (X)}

No Qiskit, podemos verificar a atuação dessas portas em um qubit genérico. 
Por simplicidade, vamos verificar a atuação da Porta $X$ nos estados da base computacional $\lbrace \vert{0}\rangle ,\vert{1}\rangle\rbrace$ como um exemplo: 
\begin{tcolorbox}[breakable, size=fbox, boxrule=1pt, pad at break*=1mm,colback=cellbackground, colframe=cellborder,coltitle=black,title=Box 8: Aplicando a Porta X]
\begin{Verbatim}[commandchars=\\\{\}]
\PY{n}{q} \PY{o}{=} \PY{n}{QuantumRegister}\PY{p}{(}\PY{l+m+mi}{1}\PY{p}{,} \PY{l+s+s1}'\PY{l+s+s1}{q}\PY{l+s+s1}'\PY{p}{)}
\PY{n}{circuito} \PY{o}{=} \PY{n}{QuantumCircuit}\PY{p}{(}\PY{n}{qubit}\PY{p}{)}
\PY{n}{estado\PYZus{}inicial} \PY{o}{=} \PY{p}{[}\PY{l+m+mi}{1}\PY{p}{,}\PY{l+m+mi}{0}\PY{p}{]}
\PY{n}{circuito}\PY{o}{.}\PY{n}{initialize}\PY{p}{(}\PY{n}{estado\PYZus{}inicial}\PY{p}{,}\PY{n}{qubit}\PY{p}{)}
\PY{n}{circuito}\PY{o}{.}\PY{n}{x}\PY{p}{(}\PY{n}{qubit}\PY{p}{)} 
\end{Verbatim}
\end{tcolorbox}

\begin{figure}[h]
    \centering
    \subfigure[]{\includegraphics[scale=0.45]{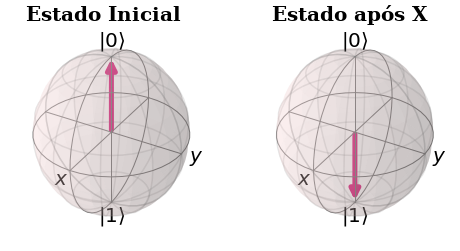}}\\
    \subfigure[]{\includegraphics[scale=0.60]{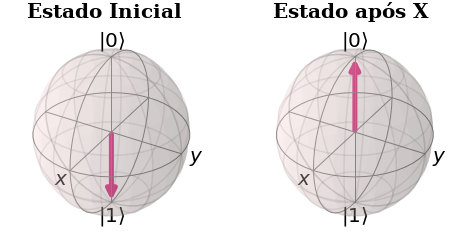}}
    \caption{Representação na esfera de Bloch da atuação da porta $X$ sobre os estados da Base Computacional (a) $\vert{0}\rangle$ (b) $\vert{1}\rangle$. Como pode ser visto, a aplicação da Porta $X$ corresponde a um inversor lógico, implementando uma operação de negação lógica.}
    \label{fig:02}
\end{figure}

A fig.~\ref{fig:02} mostra a atuação da porta X sobre os qubits da base computacional na representação da esfera de Bloch, implementado no \textit{Qiskit} conforme foi apresentado no Box 7.

Como podemos ver, a aplicação da Porta $X$ corresponde a um inversor lógico, uma vez que ela nega o valor do bit de entrada, isso pode ser interpretado como um análogo quântico para a porta \texttt{NOT}  clássica \cite{oliveira2020fisica,nielsen2002quantum}. Por esse motivo, convencionou-se chamar a porta X como Porta \texttt{NOT}  quântica  \cite{rabelo2018abordagem}.

Analogamente, para atuar as portas $Y$, $Z$ ou $S$ basta somente trocar o \texttt{x} pela letra \texttt{y}, \texttt{z} ou \texttt{s} na última linha do Box 8, respectivamente.

\textbf{Porta Hadamard (H)}

Outra porta quântica extremamente importante que atua sobre 1 qubit é a porta Hadamard ($H$).
\begin{eqnarray}
H = \frac{1}{\sqrt{2}} \left( X +Z \right)= \frac{1}{\sqrt{2}}\begin{bmatrix} 1 & 1 \\ 1 & -1 \end{bmatrix}~, \label{eq:11} 
\end{eqnarray}

Essa importância é devido ao fato da operação implementada pela porta Hadamard ser responsável pela criação de sobreposição. Considerando os estados da base computacional, a atuação da porta Hadamard resulta em: 
\begin{eqnarray}
H\vert{0}\rangle &=&  \vert{+}\rangle ~~=~~  \frac{1}{\sqrt{2}} \left( \vert{0}\rangle + \vert{1}\rangle\right)~, \label{eq:12}\\
H\vert{1}\rangle &=&  \vert{-}\rangle ~~=~~  \frac{1}{\sqrt{2}} \left( \vert{0}\rangle - \vert{1}\rangle\right)~. \label{eq:13} 
\end{eqnarray}

\begin{figure}[h]
    \centering
    \subfigure[]{\includegraphics[scale=0.45]{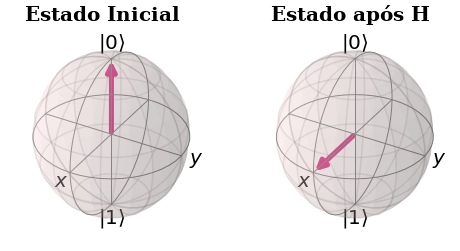}}\\
    \subfigure[]{\includegraphics[scale=0.45]{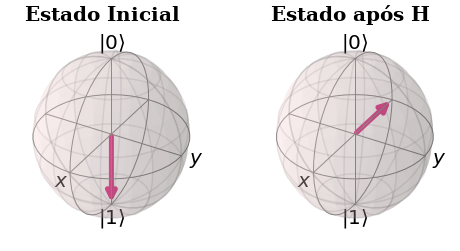}}
    \caption{Representação na esfera de Bloch da atuação da porta $H$ sobre os estados da Base Computacional (a) $\vert{0}\rangle$ (b) $\vert{1}\rangle$.}
    \label{fig:03}
\end{figure}

Analogamente a aplicação da porta \texttt{NOT}, para atuar a porta Hadamard basta somente trocar o \texttt{x} pela letra \texttt{h} na última linha do Box 8.
A fig.~\ref{fig:03} mostra a atuação da porta H sobre os qubits da base computacional na representação da esfera de Bloch.

Um fato interessante é que a porta Hadamard pode ser combinada com a porta $Z$ para formar a porta $X$, e combinada com a porta X para formar a porta Z, através das seguintes sequências:
\begin{eqnarray}
X &=&   HZH~, \label{eq:14}\\
Z &=&   HXH~. \label{eq:15} 
\end{eqnarray}

Essas combinações se mostram bastante úteis quando precisamos criar portas que não estão presentes na biblioteca do \textit{Qiskit}, como veremos a seguir na seção de aplicações.

Outro fato interessante é que as Portas de Pauli, Fase e Hadamard compõem o que chamamos de conjunto universal de portas de 1 qubit, pois através delas é possível implementar qualquer transformação unitária do estado de 1 qubit apresentado na equação~(\ref{eq:03}).

\subsubsection{Desenhando Circuitos Quânticos}

Assim como as Portas lógicas clássicas são combinadas para formar circuitos lógicos para a implementação de algoritmos, as portas lógicas quânticas também podem ser combinadas para a construção de circuitos para a implementação de algoritmos quânticos. Nesse contexto, faz-se necessário introduzir a simbologia das portas lógicas universalmente utilizadas em computação quântica. Uma vez construído o nosso circuito, podemos desenhá-lo no Jupyter notebook e exportá-lo na forma de figura, com o comando \footnote{Recomenda-se a instalação do pacote \texttt{pylatexenc} para a vizualização dos circuitos, utilizando o gerenciador de pacotes do Python (\texttt{pip}), utilizando o comando no terminal \texttt{>\ pip\ install\ pylatexenc}.}:
\begin{tcolorbox}[breakable, size=fbox, boxrule=1pt, pad at break*=1mm,colback=cellbackground, colframe=cellborder,coltitle=black,title=Box 9: Desenhando circuitos quânticos]
\begin{Verbatim}[commandchars=\\\{\}]
\PY{n}{circuito}\PY{o}{.}\PY{n}{draw}\PY{p}{(}\PY{n}{output} \PY{o}{=} \PY{l+s+s1}'\PY{l+s+s1}{mpl}\PY{l+s+s1}'\PY{p}{)}
\end{Verbatim}
\end{tcolorbox}

\begin{figure}[h]
    \centering
    \includegraphics[scale=0.55]{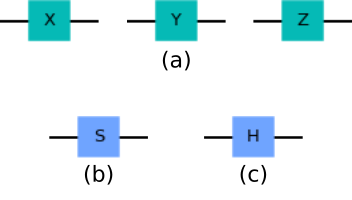}
    \caption{Representação das Portas (a) de Pauli, (b) de Fase e (c) Hadamard.}
    \label{fig:04}
\end{figure}

A fig.~\ref{fig:04} mostra a representação das portas universais de 1 qubit apresentadas nessa seção. 

\subsubsection{Portas de múltiplos qbits}

Computadores quânticos de um qubit são tão usuais quanto computadores clássicos de um bit. A grande vantagem da computação quântica aparece quando trabalhamos com sistemas de múltiplos qubits \cite{candela2015undergraduate,santos2017computador}. 

Como vimos no início dessa seção, um único bit tem dois estados possíveis $\lbrace 0,1\rbrace$, analogamente um estado qubit tem duas amplitudes complexas $\lbrace \alpha,\beta\rbrace$, equação~(\ref{eq:03}). Portanto, da mesma forma que dois bits têm quatro estados possíveis $\lbrace 00,01,10,11\rbrace$, a base computacional para um sistema de dois qubits é dada por $\lbrace \vert{00}\rangle,\vert{01}\rangle,\vert{10}\rangle,\vert{11}\rangle\rbrace$. De maneira geral, um sistema de \texttt{N}-qubits é descrito por um estado quântico
\begin{equation}
    \vert{\Psi}\rangle = c_{1} \vert{0...00}\rangle + c_{2} \vert{0...01}\rangle + c_{3} \vert{00...10}\rangle + \dots + c_{2^N} \vert{1...11}\rangle ~, \label{eq:16} 
\end{equation}
onde os \texttt{N}-qubits são considerados como um único sistema composto com $2^{N}$ estados na sua base computacional, com 
\begin{equation}
\sum\limits_{i=1}^{N} \vert c_{i}\vert^2 = 1~.
\label{eq:17}
\end{equation}

Trabalhar com vários qubits permite realizar operações em subconjuntos de qubits \cite{nielsen2002quantum} e ainda assim fazer uso das propriedades quânticas de $\vert{\Psi}\rangle$ como a sobreposição, por exemplo. Apresentamos a seguir as principais portas quânticas que operam em  múltiplos qubits usando o \textit{Qiskit}.

\textbf{Porta CNOT (CX)}

Uma das portas mais importantes de múltiplos qubits é a conhecida porta CNOT (NOT Controlado ou CX). A porta CNOT é uma porta de dois qubits, e sua atuação ocorre se, e somente se, o qubit, que chamamos de qubit de controle, for igual a $\vert{1}\rangle$. Nessa ocasião atua-se a porta NOT no estado do outro qubit, que chamamos de qubit alvo. Assim, podemos representar matricialmente a porta CNOT como:
\begin{eqnarray}
CNOT &=&  \begin{bmatrix} 1 & 0 & 0 & 0\\ 0 & 1 & 0 & 0\\ 0 & 0 & 0 & 1\\ 0 & 0 & 1 & 0\\  \end{bmatrix}~. 
\label{eq:20}
\end{eqnarray}

Considerando os estados da base computacional para dois qubits: $\lbrace \vert{00}\rangle ,\vert{01}\rangle ,\vert{10}\rangle ,\vert{11}\rangle \rbrace$, a atuação da porta CNOT resulta em: 
\begin{eqnarray}
CNOT\vert{00}\rangle &=& \vert{00}\rangle \label{eq:21}\\
CNOT\vert{01}\rangle &=& \vert{01}\rangle \label{eq:22}\\
CNOT\vert{10}\rangle &=& \vert{11}\rangle \label{eq:23}\\
CNOT\vert{11}\rangle &=& \vert{10}\rangle \label{eq:24}
\end{eqnarray}

A porta CNOT pode ser implementada em um circuito com o \texttt{qubit[0]}\footnote{Na Linguagem Python 3, o primeiro índice de uma lista é sempre o 0.} como qubit de controle e \texttt{qubit[1]} como qubit alvo da seguinte maneira: 
\begin{tcolorbox}[breakable, size=fbox, boxrule=1pt, pad at break*=1mm,colback=cellbackground, colframe=cellborder,coltitle=black,title=Box 10: Implementação da Porta CNOT]
\begin{Verbatim}[commandchars=\\\{\}]
\PY{n}{q} \PY{o}{=} \PY{n}{QuantumRegister}\PY{p}{(}\PY{l+m+mi}{2}\PY{p}{,} \PY{l+s+s1}'\PY{l+s+s1}{q}\PY{l+s+s1}'\PY{p}{)}
\PY{n}{circuito} \PY{o}{=} \PY{n}{QuantumCircuit}\PY{p}{(}\PY{n}{qubit}\PY{p}{)}
\PY{n}{circuito}\PY{o}{.}\PY{n}{cx}\PY{p}{(}\PY{n}{qubit}\PY{p}{[}\PY{l+m+mi}{0}\PY{p}{]}\PY{p}{,}\PY{n}{qubit}\PY{p}{[}\PY{l+m+mi}{1}\PY{p}{]}\PY{p}{)}
\PY{n}{circuito}\PY{o}{.}\PY{n}{draw}\PY{p}{(}\PY{n}{output} \PY{o}{=} \PY{l+s+s1}'\PY{l+s+s1}{mpl}\PY{l+s+s1}'\PY{p}{)}
\end{Verbatim}
\end{tcolorbox}

\begin{figure}[h!]
    \centering
    \includegraphics[scale=0.65]{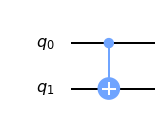}
    \caption{Representação da Porta CNOT em um circuito quântico.}
    \label{fig:05}
\end{figure}

A fig.~\ref{fig:05} mostra a representação da porta CNOT em um circuito quântico. Usando o Box 7 podemos visualizar a atuação dessa porta na representação da Esfera de Bloch, conforme pode ser visualizado na 
Fig.~\ref{fig:06}:
\begin{figure}[h]
    \centering
    \subfigure[]{\includegraphics[scale=0.25]{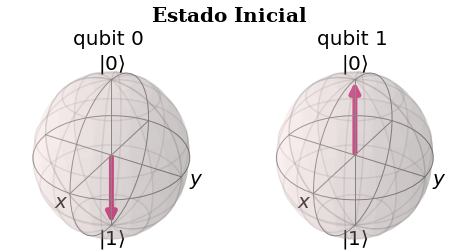}
    \includegraphics[scale=0.25]{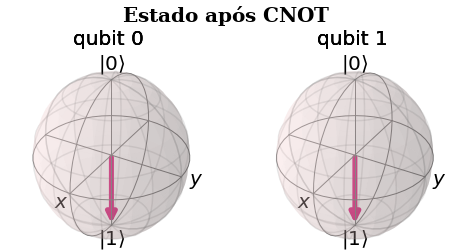}}\\
        \subfigure[]{\includegraphics[scale=0.25]{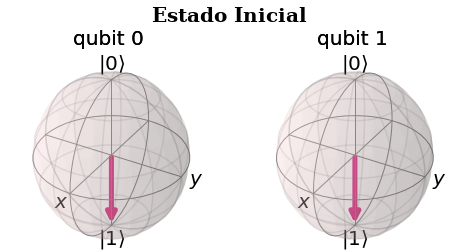}
    \includegraphics[scale=0.25]{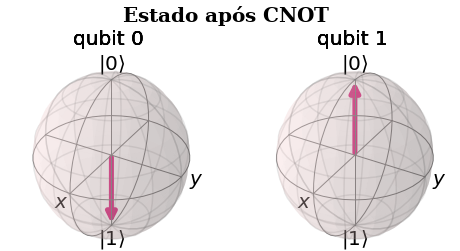}}
    \caption{Representação na esfera de Bloch da atuação da porta $CNOT$ sobre os estados da Base Computacional (a) $\vert{10}\rangle$ (b) $\vert{11}\rangle$.}
    \label{fig:06}
\end{figure}

O conjunto de portas de Pauli, de Fase, Hadamard e CNOT compõem o que chamamos de conjunto universal de Portas Lógicas quânticas \cite{oliveira2020fisica}, uma vez que é possível implementar qualquer operação unitária em um estado genérico através da combinação dessas portas.

\textbf{Porta Toffoli (CCX)}

Uma Porta de múltiplos qubits bastante presente em diversos circuitos é a porta Toffoli (CCX) \cite{nielsen2002quantum}. Sua atuação é executar a porta NOT no qubit alvo somente se dois qubits de controle estiverem no estado $\vert{1}\rangle$, sendo portanto uma porta de 3 qubits. Assim, podemos representar matricialmente a porta Toffoli como:
\begin{eqnarray}
CCX &=&  \begin{bmatrix}
1 & 0 & 0 & 0 & 0 & 0 & 0 & 0 \\
0 & 1 & 0 & 0 & 0 & 0 & 0 & 0 \\
0 & 0 & 1 & 0 & 0 & 0 & 0 & 0 \\
0 & 0 & 0 & 1 & 0 & 0 & 0 & 0 \\
0 & 0 & 0 & 0 & 1 & 0 & 0 & 0 \\
0 & 0 & 0 & 0 & 0 & 1 & 0 & 0 \\
0 & 0 & 0 & 0 & 0 & 0 & 0 & 1 \\
0 & 0 & 0 & 0 & 0 & 0 & 1 & 0 \\
\end{bmatrix}~. 
\label{eq:25}
\end{eqnarray}

A porta Toffoli pode ser implementada em um circuito com o \texttt{qubit[0]} e \texttt{qubit[1]} como qubits de controle e \texttt{qubit[2]} como qubit alvo da seguinte maneira: 
\begin{tcolorbox}[breakable, size=fbox, boxrule=1pt, pad at break*=1mm,colback=cellbackground, colframe=cellborder,coltitle=black,title=Box 11: Implementação da Porta Toffoli]
\begin{Verbatim}[commandchars=\\\{\}]
\PY{n}{q} \PY{o}{=} \PY{n}{QuantumRegister}\PY{p}{(}\PY{l+m+mi}{3}\PY{p}{,} \PY{l+s+s1}'\PY{l+s+s1}{q}\PY{l+s+s1}'\PY{p}{)}
\PY{n}{circuito} \PY{o}{=} \PY{n}{QuantumCircuit}\PY{p}{(}\PY{n}{qubit}\PY{p}{)}
\PY{n}{circuito}\PY{o}{.}\PY{n}{ccx}\PY{p}{(}\PY{n}{qubit}\PY{p}{[}\PY{l+m+mi}{0}\PY{p}{]}\PY{p}{,}\PY{n}{qubit}\PY{p}{[}\PY{l+m+mi}{1}\PY{p}{]}\PY{p}{,}\PY{n}{qubit}\PY{p}{[}\PY{l+m+mi}{2}\PY{p}{]}\PY{p}{)}
\PY{n}{circuito}\PY{o}{.}\PY{n}{draw}\PY{p}{(}\PY{n}{output} \PY{o}{=} \PY{l+s+s1}'\PY{l+s+s1}{mpl}\PY{l+s+s1}'\PY{p}{)}
\end{Verbatim}
\end{tcolorbox}

Fig.~\ref{fig:07} mostra a representação da porta Toffoli em um circuito quântico:  
\begin{figure}[h]
    \centering
    \includegraphics[scale=0.65]{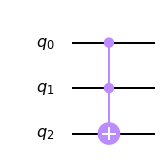}
    \caption{Representação da Porta Toffoli em um circuito quântico.}
    \label{fig:07}
\end{figure}

Novamente, podemos obter a atuação da porta Toffoli na representação da esfera de Bloch, devido à semelhança operacional com outras portas controladas como a porta CNOT. Não apresentamos essa atuação neste texto.

\subsection{Medidas e Distribuições de Probabilidade}

Como mencionamos anteriormente, definimos um conjunto de bits clássicos auxiliares de modo que as medições nos bits quânticos são salvas  como resultados clássicos $\lbrace 0,1\rbrace$. Logo, uma vez que vimos como inicializar os qubits e realizar operações universais sobre eles, temos todas as condições para implementar os nossos algoritmos quânticos. Entretanto, ainda falta um passo fundamental para completarmos nosso processo de computação: a caracterização do estado final, através distribuição de probablidade correspondente. 

Primeiramente, precisamos iniciar nosso processo computacional quântico com um estado bem definido, através das operações de inicialização discutdas anteriormente. Por simplicidade, vamos considerar o exemplo de um sistema de 2 qubits inicializados no estado $\vert{00}\rangle$ usando o comando \texttt{circuito.reset(q)}. Esse comando proporciona uma representação visual que cada qubit do sistema foi inicializado no estado  $\vert{0}\rangle$ e é bastante útil para a organização dos circuitos. Em seguida, aplicamos uma sequência de portas quânticas que manipulam os dois qubits, conforme exigido pelo algoritmo. Por exemplo, consideremos um algoritmo que coloque todos os estados da base computacional de 2 qubits $\lbrace \vert{00}\rangle ,\vert{01}\rangle ,\vert{10}\rangle ,\vert{11}\rangle \rbrace$ em um estado de sobreposição, com mesma probabilidade de medida. Isso pode ser obtido aplicando a porta Hadamard em cada qubit. O Box abaixo traz a sequência de inicialização, aplicação de portas e a realização das medidas do nosso exemplo:
\begin{tcolorbox}[breakable, size=fbox, boxrule=1pt, pad at break*=1mm,colback=cellbackground, colframe=cellborder,coltitle=black,title=Box 12: Exemplo - Criação de um estado de sobreposição para 2 qubits ]
\begin{Verbatim}[commandchars=\\\{\}]
\PY{c+c1}{\PYZsh{} Preparativos:}
\PY{n}{q} \PY{o}{=} \PY{n}{QuantumRegister}\PY{p}{(}\PY{l+m+mi}{2}\PY{p}{,} \PY{l+s+s1}'\PY{l+s+s1}{q}\PY{l+s+s1}'\PY{p}{)} \PY{c+c1}{\PYZsh{}Registrando os qubits}
\PY{n}{b} \PY{o}{=} \PY{n}{ClassicalRegister}\PY{p}{(}\PY{l+m+mi}{2}\PY{p}{,} \PY{l+s+s1}'\PY{l+s+s1}{b}\PY{l+s+s1}'\PY{p}{)} \PY{c+c1}{\PYZsh{}Registrando os Bits}
\PY{n}{circuito} \PY{o}{=} \PY{n}{QuantumCircuit}\PY{p}{(}\PY{n}{q}\PY{p}{,}\PY{n}{b}\PY{p}{)} \PY{c+c1}{\PYZsh{}Criando o Circuito}

\PY{c+c1}{\PYZsh{} Inicialização dos Estados:}
\PY{n}{circuito}\PY{o}{.}\PY{n}{reset}\PY{p}{(}\PY{n}{q}\PY{p}{)} 

\PY{c+c1}{\PYZsh{} Aplicação das Portas:}
\PY{n}{circuito}\PY{o}{.}\PY{n}{h}\PY{p}{(}\PY{n}{q}\PY{p}{)} 

\PY{c+c1}{\PYZsh{} Realização das Medidas}
\PY{n}{circuito}\PY{o}{.}\PY{n}{measure}\PY{p}{(}\PY{n}{q}\PY{p}{,}\PY{n}{b}\PY{p}{)}
\PY{n}{circuito}\PY{o}{.}\PY{n}{draw}\PY{p}{(}\PY{n}{output} \PY{o}{=} \PY{l+s+s1}'\PY{l+s+s1}{mpl}\PY{l+s+s1}'\PY{p}{)}
\end{Verbatim}
\end{tcolorbox}

\begin{figure}[h]
    \centering
    \includegraphics[scale=0.55]{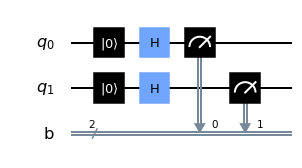}
    \caption{Circuito quântico gerador de sobreposição equiprovável para os estados da base computacional de 2 qubits.}
    \label{fig:08}
\end{figure}

A ig.~\ref{fig:08} mostra o circuito quântico gerador de sobreposição equiprovável para os estados da base computacional de 2 qubits apresentado no Box 13.  

Finalmente, após medir os qubits, podemos utilizar um computador clássico para interpretar as medições de cada qubit como resultados clássicos (0 e 1) e armazená-los nos bits clássicos definidos para esse circuito. O \textit{Qiskit} contém o simulador \texttt{QASM}, um dos principais componentes do elemento \texttt{Aer} para as simulações quânticas por meio de computação de alto desempenho. Este simulador emula a execução de circuitos quânticos em um processador local e retorna as contagens de cada medida no estado final para um dado conjunto de repetições ou \texttt{shots} do circuito definidos pelo usuário. Podemos usar esse recurso para simular  nossos circuitos quânticos usando nosso computador pessoal para emular numericamente\footnote{A descrição dos métodos numéricos utilizados nessa emulação foge dos objetivos principais deste trabalho. Para maiores detalhes sugerimos as referências \cite{qiskit-textbook,qiskit-documentation,github-qiskit,qiskit}.} um processador quântico ideal, sem nenhuma influência de pertubações externas devido ao acoplamento inevitável entre os sistemas quânticos de processamento da informação e o ambiente externo \cite{santos2017computador,ibm}, conhecido como decoerência \cite{nielsen2002quantum,oliveira2020fisica}\footnote{Na aba \textit{User Guide} do IBM QE é possível encontrar detalhes acerca do processo de decoerência que ocorre nos processadores, mas esse conteúdo encontra-se somente disponível em inglês. Para mais detalhes sobre a arquitetura do computador quântico da IBM, além disso, indicamos as referências \cite{oliveira2020fisica,santos2017computador,rabelo2018abordagem}.}.

O Box 13 a seguir apresenta o código que simula o circuito mostrado no Box 12.
\begin{tcolorbox}[breakable, size=fbox, boxrule=1pt, pad at break*=1mm,colback=cellbackground, colframe=cellborder,coltitle=black,title=Box 13: Simulando um circuito em um processador quântico ideal emulado numericamente]
\begin{Verbatim}[commandchars=\\\{\}]
\PY{n}{simular} \PY{o}{=} \PY{n}{Aer}\PY{o}{.}\PY{n}{get\PYZus{}backend}\PY{p}{(}\PY{l+s+s1}'\PY{l+s+s1}{qasm\PYZus{}simulator}\PY{l+s+s1}'\PY{p}{)}
\PY{n}{resultadolocal} \PY{o}{=} \PY{n}{execute}\PY{p}{(}\PY{n}{circuito}\PY{p}{,} \PY{n}{backend} \PY{o}{=} \PY{n}{simular}\PY{p}{,} \PY{n}{shots} \PY{o}{=} \PY{n}{8000}\PY{p}{)}\PY{o}{.}\PY{n}{result}\PY{p}{(}\PY{p}{)}
\end{Verbatim}
\end{tcolorbox}

Há pequenas variações nas probabilidades de cada estado associadas ao processo númerico que o \textit{Qiskit} utiliza para emular um processador quântico, mas há maneiras de otimizar esse processo para que a emulação se aproxime cada vez mais de um processador real. Uma delas é aumentar o número de repetições de cada circuito (\texttt{shots}), conforme apresentado no Box 13.

Assim, obtemos as contagens dos estados e obtemos a distribuição de probabilidades para cada estado. Podemos usar a função \texttt{plot{\_}histogram}, presente no módulo \texttt{qiskit.visualization}, importado no início do notebook (Box 1), para visualizar o resultado da contagem dos estados, da seguinte maneira:
\begin{tcolorbox}[breakable, size=fbox, boxrule=1pt, pad at break*=1mm,colback=cellbackground, colframe=cellborder,coltitle=black,title=Box 14: Plotando a distribuição de probabilidade do estado final de um circuito]
\begin{Verbatim}[commandchars=\\\{\}]
\PY{n}{titulo} \PY{o}{=} \PY{l+s+s1}'\PY{l+s+s1}{Probabilidades}\PY{l+s+s1}'
\PY{n}{plot\PYZus{}histogram}\PY{p}{(}\PY{n}{resultadolocal}\PY{o}{.}\PY{n}{get\PYZus{}counts}\PY{p}{(}\PY{n}{circuito}\PY{p}{)}\PY{p}{,} \PY{n}{title}\PY{o}{=}\PY{n}{titulo}\PY{p}{)}
\end{Verbatim}
\end{tcolorbox}

\begin{figure}[h]
    \centering
    \includegraphics[scale=0.45]{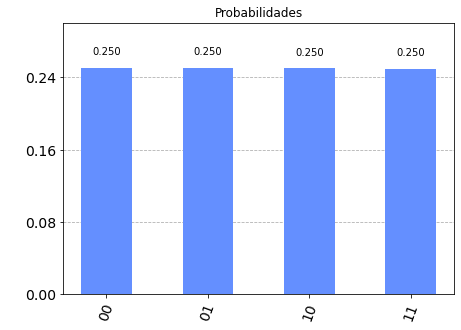}
    \caption{Distribuição de probabilidades para o circuito apresentado na Fig.~\ref{fig:08}.}
    \label{fig:09}
\end{figure}

A fig.~\ref{fig:09} mostra a distibuição de probabilidades para o estado final obido para o circuito apresentado no Box 14 (Fig.~\ref{fig:08}). Como pode ser visto, a distribuição de probabilidades obtida numericamente corresponde exatamente ao estado quântico
\begin{equation}
    \vert{\Psi}\rangle = \frac{1}{2}\left[\vert{00}\rangle +\vert{01}\rangle +\vert{10}\rangle +\vert{11}\rangle \right]~ \label{eq:16} 
\end{equation}

\subsubsection{Acessando o \textit{IBM QE} usando o \textit{Qiskit}}

Além de podermos simular nosso circuito em um processador ideal emulado numericamente, podemos também executar nossos projetos em processadores quânticos reais usando o \textit{IBM Q Experience} por meio do elemento \textit{IBM Q Provider}, que vem com o \textit{Qiskit}. Para isso, é preciso criar uma conta gratuita no \textit{IBM Q Experience} \cite{ibm}. Acessando as configurações de \textit{Minha Conta}, o usuário encontra seu token de API, que é necessario para acessar dispositivos IBM Q de seu computador doméstico usando o \textit{Qiskit}. No Notebook Jupyter, podemos usar os seguintes comandos:
\begin{tcolorbox}[breakable, size=fbox, boxrule=1pt, pad at break*=1mm,colback=cellbackground, colframe=cellborder,coltitle=black,title=Box 15: Salvando a conta IBM QE no computador]
\begin{Verbatim}[commandchars=\\\{\}]
\PY{n}{IBMQ}\PY{o}{.}\PY{n}{save\PYZus{}account}\PY{p}{(}\PY{l+s+s1}'\PY{l+s+s1}{Users\PYZus{}Token}\PY{l+s+s1}'\PY{p}{)}
\end{Verbatim}
\end{tcolorbox}

Este comando irá salvar o token de API do usuário em seu computador, permitindo acessar dispositivos quânticos disponibilizados pela IBM. Esse passo só precisa ser realizado uma vez. 

Para carregar a conta, usamos o comando:
\begin{tcolorbox}[breakable, size=fbox, boxrule=1pt, pad at break*=1mm,colback=cellbackground, colframe=cellborder,coltitle=black,title=Box 16: Carregando a conta IBM QE no notebook do Jupyter]
\begin{Verbatim}[commandchars=\\\{\}]
\PY{n}{IBMQ}\PY{o}{.}\PY{n}{load\PYZus{}account}\PY{p}{(}\PY{p}{)}
\end{Verbatim}
\end{tcolorbox}

Após o comando apresentado no Box 16 ser executado, a conta será carregada com êxito e poderemos ver a conta de acesso através da saída:
\begin{tcolorbox}[breakable, size=fbox, boxrule=1pt, pad at break*=1mm,colback=cellbackground, colframe=cellborder,coltitle=black,title=Box 17: Saída padrão atestando o acesso aos hardwares disponibilizados pela IBM]
\begin{Verbatim}[commandchars=\\\{\}]
<AccountProvider for IBMQ(hub='ibm-q', group='open', project='main')>
\end{Verbatim}
\end{tcolorbox}

Ao concluir esta etapa, podemos executar nossos projetos, não apenas em um processador emulado em um computador doméstico, mas também enviar circuitos quânticos para dispositivos da IBM e obter os resultados em hardware quântico real. Vejamos o exemplo do circuito gerador de sobreposição equiprovável para os estados da base computacional de 2 qubits, apresentado na fig.~\ref{fig:08}.

O Box 18 lista os comandos que selecionam o provedor e os sistemas quânticos e simuladores aos quais temos acesso pelo IBM QE para realizarmos a computação em um processadores reais\footnote{Há 9 sistemas disponíveis para a execução. Durante o \textit{paper} usamos sempre os \textit{hardwares} que estavam com a menor quantidade de trabalhos em execução, o que pode ser conferido na painel de controle do IBM QE \cite{ibm}. Todos os algoritmos apresentados nesse trabalho foram executados no mesmo processador quântico de 5 qubits \texttt{ibmq{\_}valencia}, conforme mostramos no Box 18.}. 
 \begin{tcolorbox}[breakable, size=fbox, boxrule=1pt, pad at break*=1mm,colback=cellbackground, colframe=cellborder,coltitle=black,title=Box 18: Selecionando o provedor e executando o trabalho]
\begin{Verbatim}[commandchars=\\\{\}]
\PY{n}{provedor} \PY{o}{=} \PY{n}{IBMQ}\PY{o}{.}\PY{n}{get\PYZus{}provider}\PY{p}{(}\PY{l+s+s1}'\PY{l+s+s1}{ibm\PYZhy{}q}\PY{l+s+s1}'\PY{p}{)}
\PY{n}{comput} \PY{o}{=} \PY{n}{provider}\PY{o}{.}\PY{n}{get\PYZus{}backend}\PY{p}{(}\PY{l+s+s1}'\PY{l+s+s1}{ibmq\PYZus{}valencia}\PY{l+s+s1}'\PY{p}{)}
\PY{n}{trabalho} \PY{o}{=} \PY{n}{execute}\PY{p}{(}\PY{n}{circuito}\PY{p}{,} \PY{n}{backend}\PY{o}{=}\PY{n}{comput}\PY{p}{,} \PY{n}{shots} \PY{o}{=} \PY{n}{8000}\PY{p}{)}
\PY{n}{job\PYZus{}monitor}\PY{p}{(}\PY{n}{trabalho}\PY{p}{)}
\end{Verbatim}
\end{tcolorbox}

Usando o comando \texttt{job{\_}monitor()} podemos monitorar o nosso circuito na fila de execução do processador em tempo real. Após finalizar a execução, recebemos a mensagem (em inglês): 

    \begin{Verbatim}[commandchars=\\\{\}]
Job Status: job has successfully run
    \end{Verbatim}

Indicando que o trabalho foi executado com sucesso. Assim, obtemos a contagem dos estados e podemos plotar as distribuições de probabilidade como apresentado no Box 15\footnote{Devido a limitações de hardware, o número de repetições permitidas (\texttt{shots}) no processador quântico real era de 8192.}. Usando os comandos apresentados no Box 19 (a seguir), fazemos o comparativo das distribuições de probabilidades obtidas para o circuito da fig~\ref{fig:08}, simulado numericamente em um computador doméstico e executado em um processador quântico real. 
\begin{tcolorbox}[breakable, size=fbox, boxrule=1pt, pad at break*=1mm,colback=cellbackground, colframe=cellborder,coltitle=black,title=Box 19: Plotando as distribuições de probabilidade de um circuito emulado numericamente e em um processador real.]
\begin{Verbatim}[commandchars=\\\{\}]
\PY{n}{resultadoIBM} \PY{o}{=} \PY{n}{trabalho}\PY{o}{.}\PY{n}{result}\PY{p}{(}\PY{p}{)}
\PY{n}{legenda} \PY{o}{=} \PY{p}{[}\PY{l+s+s1}'\PY{l+s+s1}{Processador Ideal}\PY{l+s+s1}'\PY{p}{,} \PY{l+s+s1}'\PY{l+s+s1}{Processador Real}\PY{l+s+s1}'\PY{p}{]}
\PY{n}{titulo} \PY{o}{=} \PY{l+s+s1}'\PY{l+s+s1}{Probabilidades}\PY{l+s+s1}'
\PY{n}{plot\PYZus{}histogram}\PY{p}{(}\PY{p}{[}\PY{n}{resultadolocal}\PY{o}{.}\PY{n}{get\PYZus{}counts}\PY{p}{(}\PY{n}{circuito}\PY{p}{)}\PY{p}{,} \PY{n}{resultadoIBM}\PY{o}{.}\PY{n}{get\PYZus{}counts}\PY{p}{(}\PY{n}{circuito}\PY{p}{)}\PY{p}{]}\PY{p}{,} \PY{n}{legend}\PY{o}{=}\PY{n}{legenda}\PY{p}{,}\PY{n}{title}\PY{o}{=}\PY{n}{titulo}\PY{p}{)}
\end{Verbatim}
\end{tcolorbox}

\begin{figure}[h]
    \centering
    \includegraphics[scale=0.45]{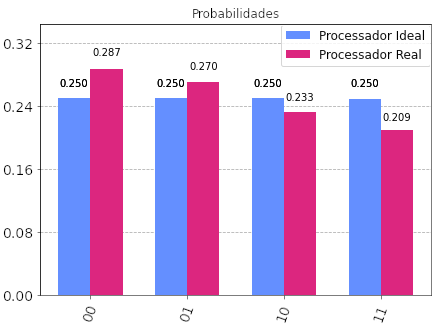}
    \caption{Distribuição de probabilidade para o circuito apresentado na fig.~\ref{fig:08} emulado numericamente (processador ideal) e em um processador real.}
    \label{fig:010}
\end{figure}

Fica clara a diferença entre um processador quântico ideal (emulado em um computador doméstico) e um processador quântico real, devido aos efeitos de decoerência. As referências \cite{santos2017computador,rabelo2018abordagem} trazem uma análise detalhada do processador quântico de 5 qubits da IBM QE, cuja a arquitetura é a mesma do utilizado nesse trabalho (\texttt{ibmq{\_}valencia}). Uma descrição da análise de dados \cite{santos2017computador,rabelo2018abordagem} e o erro padrão associado a essa arquitetura de processador quântico \cite{rabelo2018abordagem}.

\subsection{Emaranhamento Quântico}

Uma vez que conhecemos como inicializar nossos qubits e vimos as principais operações que atuam sobre eles, podemos introduzir uma das principais propriedades da mecânica quântica e um recurso fundamental para o processamento da informação quântica, o \textit{Emaranhamento}.

O emaranhamento quântico é um dos fenômenos mais interessantes da mecânica quântica que emerge da interação entre múltiplos qubits. Hoje em dia, o emaranhamento quântico tem recebido atenção considerável como um recurso notável para o processamento de informação quântica \cite{nielsen2002quantum,horodecki} e para a compreensão de correlações em sistemas compostos. Einstein, Podolsky, e Rosen (EPR) introduziram a ideia de que estados quânticos de um sistema composto podem apresentar correlações não locais entre seus componentes. Schr\"odinger analisou algumas consequências físicas da mecânica quântica, observando que alguns estados quânticos bipartidos (estados EPR \cite{horodecki}) não admitiam atribuir estados individuais de subsistemas, implicando em algumas \textit{predições emaranhadas} para a natureza quântica dos subsistemas \cite{horodecki}. Portanto, o emaranhamento implica a existência de estados quânticos globais de sistemas compostos que não podem ser escritos como um produto dos estados quânticos de subsistemas individuais \cite{nielsen2002quantum,horodecki}.

Consideremos um estado quântico de um sistema composto perfeitamente descrito pela função de onda
\begin{equation}
\vert \Psi\rangle\neq\vert \phi_1\rangle\otimes\vert \phi_2\rangle\otimes\cdots\otimes\vert \phi_n\rangle~,
\label{eq:302}
\end{equation}
não podemos especificar qualquer estado quântico puro $\vert \phi_i\rangle$ ($i=1,\dots ,n$) dos subsistemas separadamente; isto é, o conhecimento de um todo não implica conhecimento das partes.

Portanto, não sabemos nada sobre os subsistemas, embora tenhamos conhecimento do sistema como um todo, uma vez que conhecemos $\vert \Psi\rangle$. Isso contrasta com a situação clássica, em que sempre podemos considerar os estados individuais dos subsistemas. Esta é uma pista de que estados emaranhados são estados correlacionados especiais, cuja natureza física não pode ser simulada ou representada a partir de correlações clássicas. 

Em um circuito quântico, podemos emaranhar dois qubits através da combinação das portas Hadamard e CNOT, apresentadas anteriormente.
Dependendo dos valores de inicialização dos qubits \texttt{q[0]} e \texttt{q[1]}, obtemos um dos quatro estados maximamente emaranhados para 2 qubits ou \textit{Estados de Bell} \cite{oliveira2020fisica}. 
\begin{table}[h]
\caption{Estados Maximamente Emaranhados correspondente a cada inicialização dos qubits \texttt{q[0]} e \texttt{q[1].}}
\centering
\begin{tabular}{|c|c|c|}
\hline
\texttt{q[0]} & \texttt{q[1]} & $\vert{\Psi}\rangle$ \\ \hline
$\vert{0}\rangle$ & $\vert{0}\rangle$ & $\frac{1}{\sqrt{2}}\left[\vert{00}\rangle +\vert{11}\rangle \right]$ \\ \hline
$\vert{0}\rangle$ & $\vert{1}\rangle$ & $\frac{1}{\sqrt{2}}\left[\vert{01}\rangle +\vert{10}\rangle \right]$ \\ \hline
$\vert{1}\rangle$ & $\vert{0}\rangle$ & $\frac{1}{\sqrt{2}}\left[\vert{00}\rangle -\vert{11}\rangle \right]$ \\ \hline
$\vert{1}\rangle$ & $\vert{1}\rangle$ &$\frac{1}{\sqrt{2}}\left[\vert{01}\rangle +\vert{10}\rangle \right]$ \\ \hline
\end{tabular}
\label{tab:01}
\end{table}

Vejamos um exemplo para o sistema inicializado no estado $\texttt{q[0]} = \vert{0}\rangle$ e $\texttt{q[1]} = \vert{0}\rangle$. Ao final do processo executado em um \textit{hardware} quântico real, realizamos medidas para um conjunto de repetições e obtemos a distribuição de probabilidade correspondente ao estado final.
\begin{tcolorbox}[breakable, size=fbox, boxrule=1pt, pad at break*=1mm,colback=cellbackground, colframe=cellborder,coltitle=black,title=Box 20: Circuito quântico gerador de estados quânticos emaranhado para 2 qubits ]
\begin{Verbatim}[commandchars=\\\{\}]
\PY{c+c1}{\PYZsh{} Preparativos:}

\PY{n}{q} \PY{o}{=} \PY{n}{QuantumRegister}\PY{p}{(}\PY{l+m+mi}{2}\PY{p}{,} \PY{l+s+s1}'\PY{l+s+s1}{q}\PY{l+s+s1}'\PY{p}{)} \PY{c+c1}{\PYZsh{}Registrando os qubits}
\PY{n}{b} \PY{o}{=} \PY{n}{ClassicalRegister}\PY{p}{(}\PY{l+m+mi}{2}\PY{p}{,} \PY{l+s+s1}'\PY{l+s+s1}{b}\PY{l+s+s1}'\PY{p}{)} \PY{c+c1}{\PYZsh{}Registrando os Bits}
\PY{n}{circuito} \PY{o}{=} \PY{n}{QuantumCircuit}\PY{p}{(}\PY{n}{q}\PY{p}{,}\PY{n}{b}\PY{p}{)} \PY{c+c1}{\PYZsh{}Criando o Circuito}

\PY{c+c1}{\PYZsh{} Inicialização dos Estados:}
\PY{n}{circuito}\PY{o}{.}\PY{n}{reset}\PY{p}{(}\PY{n}{q}\PY{p}{[}\PY{l+m+mi}{0}\PY{p}{]}\PY{p}{)} 
\PY{n}{circuito}\PY{o}{.}\PY{n}{reset}\PY{p}{(}\PY{n}{q}\PY{p}{[}\PY{l+m+mi}{1}\PY{p}{]}\PY{p}{)}

\PY{c+c1}{\PYZsh{} Aplicação das Portas:}
\PY{n}{circuito}\PY{o}{.}\PY{n}{h}\PY{p}{(}\PY{n}{q}\PY{p}{[}\PY{l+m+mi}{0}\PY{p}{]}\PY{p}{)} 
\PY{n}{circuito}\PY{o}{.}\PY{n}{cx}\PY{p}{(}\PY{n}{q}\PY{p}{[}\PY{l+m+mi}{0}\PY{p}{]}\PY{p}{,}\PY{n}{q}\PY{p}{[}\PY{l+m+mi}{1}\PY{p}{]}\PY{p}{)} 

\PY{c+c1}{\PYZsh{} Realização das Medidas}
\PY{n}{circuito}\PY{o}{.}\PY{n}{measure}\PY{p}{(}\PY{n}{q}\PY{p}{,}\PY{n}{b}\PY{p}{)}
\PY{n}{circuito}\PY{o}{.}\PY{n}{draw}\PY{p}{(}\PY{n}{output} \PY{o}{=} \PY{l+s+s1}'\PY{l+s+s1}{mpl}\PY{l+s+s1}'\PY{p}{)}
\end{Verbatim}
\end{tcolorbox}
 
A fig.~\ref{fig:11} mostra a representação do circuito representado no Box 20 e a distribuição de probabilidades correspondente ao resultado desse circuito executado em um hardware quântico real, utilizando o IBM QE:
\begin{figure}[h]
    \centering
    \subfigure[]{\includegraphics[scale=0.55]{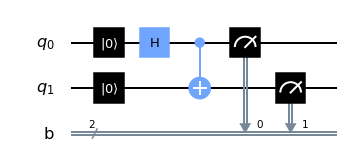}}
    \subfigure[]{\includegraphics[scale=0.45]{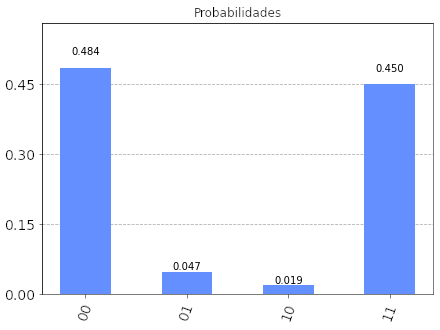}}
    \caption{(a) Representação do circuito gerador de estados quânticos emaranhado para 2 qubits. (b) Distribuição de probabilidade correspondente ao resultado desse circuito executado em um \textit{hardware} quântico real usando o IBM QE, com os qubits inicializados no estado $\texttt{q[0]} = \vert{0}\rangle$ e $\texttt{q[1]} = \vert{0}\rangle$.} 
    \label{fig:11}
\end{figure}

\section{Aplicações}
\label{application}

Uma vez que sabemos inicializar os qubits, emaranhá-los, aplicar as portas quânticas e obter os resultados através de medidas, temos todas as condições de construir algoritmos para a solução de problemas quânticos simples. Nessa seção, traremos algumas aplicações de algorimos quânticos executados em computadores quânticos reais. Apresentaremos a construção de portas lógicas clássicas a partir de portas quânticas, o famoso algoritmo de teleporte quântico \cite{santos2017computador,rabelo2018abordagem,nielsen2002quantum,oliveira2020fisica} e o algoritmo de busca de Grover \cite{nielsen2002quantum,castillo2020classical}. Os códigos são apresentados ao longo do texto, de modo que os leitores possam reproduzi-los em seus computadores, podendo inclusive construir os seus próprios projetos a partir deles.

\subsection{Simulando portas lógicas clássicas usando portas quânticas}

Uma porta lógica clássica pode ser definida como um modelo ou dispositivo físico que implementa uma determinada função booleana \cite{whitesitt2012boolean}, realizando assim aquilo que é conhecido como operação lógica. Essa operação é realizada em uma (porta \texttt{NOT}, por exemplo) ou mais entradas binárias  (bits), produzindo somente uma única saída $\lbrace 0,1\rbrace$.

Existe um conjunto de portas lógicas (clássicas) a partir das quais podemos construir qualquer operação computacional em um computador clássico \cite{oliveira2020fisica}.  Essas são as portas \texttt{AND},  \texttt{OR} e \texttt{NOT}, também conhecidas como conjunto de portas universais da Álgebra Booleana. 

Como vimos anteriormente, a porta quântica $X$ corresponde ao análogo quântico da porta \texttt{NOT} clássica. A seguir, apresentamos como podemos construir as portas \texttt{AND} e \texttt{OR}, e os seus resultados executados em um computador quântico real.

\subsubsection{Porta AND}

A porta clássica \texttt{AND} implementa o que chamamos de conjunção lógica \cite{whitesitt2012boolean}. A tabela \ref{tab:02} traz o que chamamos de \textit{tabela verdade} para essa operação lógica, a partir da qual é possível  definir o resultado lógico dessa operação. 

\begin{table}[h]
\caption{Tabela verdade para a porta lógica clássica AND}
\centering
\begin{tabular}{|c|c|c|}
\hline
\multicolumn{2}{|c|}{Entrada} & Saída \\ \hline
\texttt{q[0]} & \texttt{q[1]} & \texttt{q[2]}\\ \hline
0 & 0 & 0 \\ \hline
0 & 1 & 0 \\ \hline
1 & 0 & 0 \\ \hline
1 & 1 & 1 \\ \hline
\end{tabular}
\label{tab:02}
\end{table}
Como pode ser visto, a partir de dois bits de entrada, a saída \texttt{1} é obtida somente se as duas entradas também forem 1. Assim, podemos dizer que a porta \texttt{AND} encontra o valor mínimo entre dois bits.

Quanticamente, a porta \texttt{AND} pode ser implementada a partir da porta Toffoli, conforme apresentamos no Box 11. Como todas as portas clássicas, exceto a porta \texttt{NOT}, a porta \texttt{AND} não é reversível. Entretanto, como toda porta quântica, a porta Toffoli é reversível, o que significa que implementar a porta \texttt{AND} em computadores quânticos permite a construção de circuitos reversíveis. O Box 22 apresenta a construção do circuito quântico para a porta AND. Como as portas clássicas têm somente uma saída, a medida é realizada apenas no qubit alvo da porta Toffoli.

\begin{tcolorbox}[breakable, size=fbox, boxrule=1pt, pad at break*=1mm,colback=cellbackground, colframe=cellborder, coltitle=black, title=Box 21: Criando o circuito para a porta \texttt{AND}]
\begin{Verbatim}[commandchars=\\\{\}]
\PY{n}{q} \PY{o}{=} \PY{n}{QuantumRegister}\PY{p}{(}\PY{l+m+mi}{3}\PY{p}{,} \PY{l+s+s1}'\PY{l+s+s1}{q}\PY{l+s+s1}'\PY{p}{)}
\PY{n}{b} \PY{o}{=} \PY{n}{ClassicalRegister}\PY{p}{(}\PY{l+m+mi}{1}\PY{p}{,} \PY{l+s+s1}'\PY{l+s+s1}{b}\PY{l+s+s1}'\PY{p}{)}
\PY{n}{circuito} \PY{o}{=} \PY{n}{QuantumCircuit}\PY{p}{(}\PY{n}{q}\PY{p}{,} \PY{n}{b}\PY{p}{)} \PY{c+c1}
\PY{n}{circuito}\PY{o}{.}\PY{n}{ccx}\PY{p}{(}\PY{n}{q}\PY{p}{[}\PY{l+m+mi}{0}\PY{p}{]}\PY{p}{,}\PY{n}{q}\PY{p}{[}\PY{l+m+mi}{1}\PY{p}{]}\PY{p}{,}\PY{n}{q}\PY{p}{[}\PY{l+m+mi}{2}\PY{p}{]}\PY{p}{)}
\PY{n}{circuito}\PY{o}{.}\PY{n}{measure}\PY{p}{(}\PY{n}{q}\PY{p}{[}\PY{l+m+mi}{2}\PY{p}{]}\PY{p}{,} \PY{n}{b}\PY{p}{)}
\PY{n}{circuito}\PY{o}{.}\PY{n}{draw}\PY{p}{(}\PY{n}{output} \PY{o}{=} \PY{l+s+s1}'\PY{l+s+s1}{mpl}\PY{l+s+s1}'\PY{p}{)}
\end{Verbatim}
\end{tcolorbox}

\begin{figure}[h]
    \centering
    \includegraphics[scale=0.55]{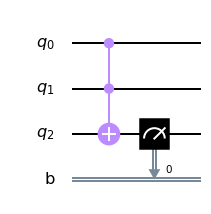}
    \caption{Representação do circuito para a aplicação da porta and.} 
    \label{fig:12}
\end{figure}

A fig.~\ref{fig:12} apresenta o circuito quântico que implementa a porta AND.

\begin{figure}[h]
    \centering
    \includegraphics[scale=0.45]{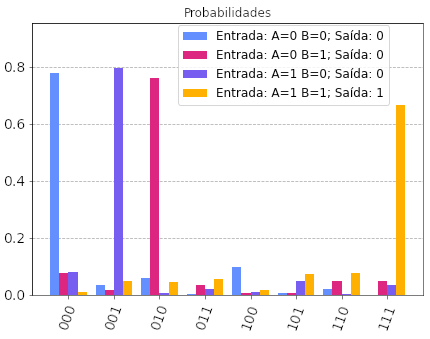}
    \caption{Distrituição de probabilidade para o circuito apresentado na figura 12 executado em um computador quântico real. Apresentamos os resultados da tabela verdade correspondente à porta clássica \texttt{AND} (Tabela \ref{tab:02}).} 
    \label{fig:13}
\end{figure}

A distribuição de probabilidade correspondente à aplicação desse circuito em um processador quântico real é apresentada na fig. \ref{fig:13}\footnote{Nesse ponto, vale destacar que a apresentação dos bits no eixo \textit{x} em todas as distribuições de probabilidade desse artigo seguem o padrão do \textit{Qiskit}: \texttt{bit[0]bit[1]bit[2]} apresentados de cima para baixo.}. Apresentamos os resultados correspondentes à tabela verdade da porta \texttt{AND} clássica (Tabela \ref{tab:02}). Como pode ser visto, a probabilidade não é de 100\% para os estados esperados devido aos efeitos de decoerência do processador real, conforme discutido anteriormente.

\subsubsection{Porta OR}

A porta \texttt{OR} é uma porta clássica universal que implementa o que chamamos em álgebra booleana de disjunção lógica \cite{whitesitt2012boolean}. A tabela verdade para a aplicação da porta clássica \texttt{OR} é apresentada na Tabela \ref{tab:03}.

\begin{table}[h]
\caption{Tabela verdade para a porta lógica clássica OR.}
\centering
\begin{tabular}{|c|c|c|}
\hline
\multicolumn{2}{|c|}{Entrada} & Saída \\ \hline
\texttt{q[0]} & \texttt{q[1]} & \texttt{q[2]}\\ \hline
0 & 0 & 0 \\ \hline
0 & 1 & 1 \\ \hline
1 & 0 & 1 \\ \hline
1 & 1 & 1 \\ \hline
\end{tabular}
\label{tab:03}
\end{table}

Como pode ser visto, uma saída \texttt{1} é obtida se pelo menos uma das entradas for \texttt{1}. Assim, dizemos que a porta \texttt{OR} encontra o máximo entre duas entradas binárias.

O análogo quântico para a \texttt{OR} pode ser construído através da combinação das portas Toffoli e CNOT. O Box 22 traz o código de construção do circuito para a implementação da porta \texttt{OR}.
\begin{tcolorbox}[breakable, size=fbox, boxrule=1pt, pad at break*=1mm,colback=cellbackground, colframe=cellborder, coltitle=black, title=Box 22: Criando o circuito para a porta \texttt{OR}]
\begin{Verbatim}[commandchars=\\\{\}]
\PY{n}{q} \PY{o}{=} \PY{n}{QuantumRegister}\PY{p}{(}\PY{l+m+mi}{3}\PY{p}{,} \PY{l+s+s1}'\PY{l+s+s1}{q}\PY{l+s+s1}'\PY{p}{)}
\PY{n}{b} \PY{o}{=} \PY{n}{ClassicalRegister}\PY{p}{(}\PY{l+m+mi}{1}\PY{p}{,} \PY{l+s+s1}'\PY{l+s+s1}{b}\PY{l+s+s1}'\PY{p}{)}
\PY{n}{circuito} \PY{o}{=} \PY{n}{QuantumCircuit}\PY{p}{(}\PY{n}{q}\PY{p}{,} \PY{n}{b}\PY{p}{)} \PY{c+c1}
\PY{n}{circuito}\PY{o}{.}\PY{n}{cx}\PY{p}{(}\PY{n}{q}\PY{p}{[}\PY{l+m+mi}{1}\PY{p}{]}\PY{p}{,}\PY{n}{q}\PY{p}{[}\PY{l+m+mi}{2}\PY{p}{]}\PY{p}{)}
\PY{n}{circuito}\PY{o}{.}\PY{n}{cx}\PY{p}{(}\PY{n}{q}\PY{p}{[}\PY{l+m+mi}{0}\PY{p}{]}\PY{p}{,}\PY{n}{q}\PY{p}{[}\PY{l+m+mi}{2}\PY{p}{]}\PY{p}{)}
\PY{n}{circuito}\PY{o}{.}\PY{n}{ccx}\PY{p}{(}\PY{n}{q}\PY{p}{[}\PY{l+m+mi}{0}\PY{p}{]}\PY{p}{,}\PY{n}{q}\PY{p}{[}\PY{l+m+mi}{1}\PY{p}{]}\PY{p}{,}\PY{n}{q}\PY{p}{[}\PY{l+m+mi}{2}\PY{p}{]}\PY{p}{)}
\PY{n}{circuito}\PY{o}{.}\PY{n}{measure}\PY{p}{(}\PY{n}{q}\PY{p}{[}\PY{l+m+mi}{2}\PY{p}{]}\PY{p}{,} \PY{n}{b}\PY{p}{)}
\PY{n}{circuito}\PY{o}{.}\PY{n}{draw}\PY{p}{(}\PY{n}{output} \PY{o}{=} \PY{l+s+s1}'\PY{l+s+s1}{mpl}\PY{l+s+s1}'\PY{p}{)}
\end{Verbatim}
\end{tcolorbox}

\begin{figure}[h]
    \centering
    \includegraphics[scale=0.55]{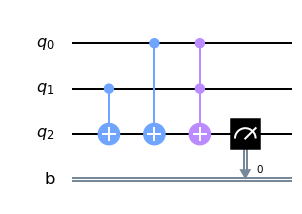}
    \caption{Representação do circuito para a aplicação da porta OR.} 
    \label{fig:14}
\end{figure}

O circuito para essa operação é apresentado na figura \ref{fig:14}.

\begin{figure}[h]
    \centering
    \includegraphics[scale=0.45]{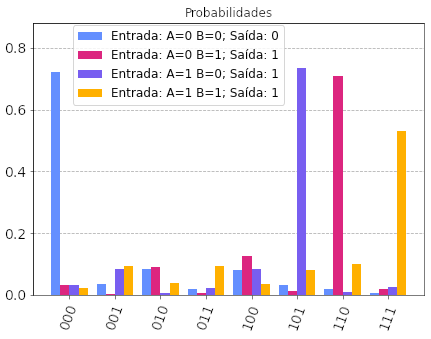}
    \caption{Distrituição de probabilidade para o circuito apresentado na figura \ref{fig:14} executado em um computador quântico real. Apresentamos os resultados da tabela verdade correspondente a porta classica \texttt{OR} (Tabela \ref{tab:03}).} 
    \label{fig:15}
\end{figure}

A distribuição de probabilidade correspondente à tabela verdade da porta OR (Tabela \ref{tab:03}), executada em um processador quântico real é apresentada na fig. \ref{fig:15}.

Assim, através das portas \texttt{NOT}  (fig. \ref{fig:02}), \texttt{AND} (fig. \ref{fig:13}) e \texttt{OR} (fig. \ref{fig:15}) é possivel implementar qualquer porta lógica clássica em um computador quântico, com a vantagem de que as portas \texttt{AND} e \texttt{OR} quânticas são reversíveis, ao contrário de seus análogos clássicos \cite{oliveira2020fisica}.

\subsection{Teleporte Quântico}

Outra aplicação muito interessante e bastante discutida na literatura da computação quântica é o \textit{Teleporte Quântico} \cite{nielsen2002quantum,oliveira2020fisica,santos2017computador,rabelo2018abordagem,bennett2000quantum}. O protocolo de teleporte quântico consiste em uma operação de transmissão de um estado quântico entre duas partes, convencionalmente conhecidas como Alice e Bob, separadas espacialmente \cite{nielsen2002quantum,oliveira2020fisica,santos2017computador,rabelo2018abordagem,bennett2000quantum}, usando dois qubits emaranhados. Assim, podemos imaginar que Alice queira enviar um estado quântico puro qualquer para Bob. Para isso, Alice deverá preparar o qubit que cuja informação será enviada, e possuir um segundo qubit que será emaranhado ao qubit de Bob, que receberá a informação. 

Vamos considerar em nosso exemplo que Alice pretende teleportar o estado:
\begin{equation}
    \vert{\psi}\rangle = \sqrt{\frac{1}{3}}\vert{0}\rangle + \sqrt{\frac{2}{3}}\vert{1}\rangle~,
    \label{estadoalice}
\end{equation}
para isso iremos inicializar o circuito o qubit de Alice nesse estado. O Box a seguir apresenta o registro dos qubits de Alice e Bob, a inicialização do estado que será teleportado  e o registro do bit clássico que Bob armazena o resultado da medida no seu estado recebido.
\begin{tcolorbox}[breakable, size=fbox, boxrule=1pt, pad at break*=1mm, colback=cellbackground, colframe=cellborder, coltitle=black, title=Box 23: Registrando os qubits e inicializando o estado que será teleportado]
\begin{Verbatim}[commandchars=\\\{\}]
\PY{n}{Alice} \PY{o}{=} \PY{n}{QuantumRegister}\PY{p}{(}\PY{l+m+mi}{2}\PY{p}{,} \PY{l+s+s1}'\PY{l+s+s1}{alice}\PY{l+s+s1}'\PY{p}{)}
\PY{n}{ba} \PY{o}{=} \PY{n}{ClassicalRegister}\PY{p}{(}\PY{l+m+mi}{1}\PY{p}{,} \PY{l+s+s1}'\PY{l+s+s1}{c\PYZus{}alice}\PY{l+s+s1}'\PY{p}{)}
\PY{n}{Bob} \PY{o}{=} \PY{n}{QuantumRegister}\PY{p}{(}\PY{l+m+mi}{1}\PY{p}{,} \PY{l+s+s1}'\PY{l+s+s1}{bob}\PY{l+s+s1}'\PY{p}{)}
\PY{n}{b} \PY{o}{=} \PY{n}{ClassicalRegister}\PY{p}{(}\PY{l+m+mi}{1}\PY{p}{,} \PY{l+s+s1}'\PY{l+s+s1}{c\PYZus{}bob}\PY{l+s+s1}'\PY{p}{)}
\PY{n}{teleporte} \PY{o}{=} \PY{n}{QuantumCircuit}\PY{p}{(}\PY{n}{Alice}\PY{p}{,}\PY{n}{Bob}\PY{p}{,}\PY{n}{b}\PY{p}{)}
\PY{n}{estado\PYZus{}inicial} \PY{o}{=} \PY{p}{[}\PY{n}{np}\PY{o}{.}\PY{n}{sqrt}\PY{p}{(}\PY{l+m+mi}{1}\PY{o}{/}\PY{l+m+mi}{3}\PY{p}{)}\PY{p}{,}\PY{n}{np}\PY{o}{.}\PY{n}{sqrt}\PY{p}{(}\PY{l+m+mi}{2}\PY{o}{/}\PY{l+m+mi}{3}\PY{p}{)}\PY{p}{]}
\PY{n}{teleporte}\PY{o}{.}\PY{n}{initialize}\PY{p}{(}\PY{n}{estado\PYZus{}inicial}\PY{p}{,}\PY{n}{Alice}\PY{p}{[}\PY{l+m+mi}{0}\PY{p}{]}\PY{p}{)}
\PY{n}{teleporte}\PY{o}{.}\PY{n}{barrier}\PY{p}{(}\PY{p}{)}
\end{Verbatim}
\end{tcolorbox}

O próximo passo é emaranhar o qubit auxiliar de Alice com o qubit de Bob em um dos Estados de Bell apresentados na Tabela \ref{tab:01}, usando o circuito quântico gerador de estados quânticos emaranhado para 2 qubits (Box 21). O Box 24 apresenta o circuito gerador de emaranhamento entre o qubit de Alice e Bob:
\begin{tcolorbox}[breakable, size=fbox, boxrule=1pt, pad at break*=1mm, colback=cellbackground, colframe=cellborder, coltitle=black, title=Box 24: Emaranhando o qubit auxiliar de Alice com o qubit de Bob]
\begin{Verbatim}[commandchars=\\\{\}]
\PY{n}{teleporte}\PY{o}{.}\PY{n}{h}\PY{p}{(}\PY{n}{Bob}\PY{p}{[}\PY{l+m+mi}{0}\PY{p}{]}\PY{p}{)}
\PY{n}{teleporte}\PY{o}{.}\PY{n}{cx}\PY{p}{(}\PY{n}{Bob}\PY{p}{[}\PY{l+m+mi}{0}\PY{p}{]}\PY{p}{,}\PY{n}{Alice}\PY{p}{[}\PY{l+m+mi}{1}\PY{p}{]}\PY{p}{)}
\PY{n}{teleporte}\PY{o}{.}\PY{n}{barrier}\PY{p}{(}\PY{p}{)}
\end{Verbatim}
\end{tcolorbox}

Em seguida, Alice inicia o processo de envio do estado preparado, no Box 25.
\begin{tcolorbox}[breakable, size=fbox, boxrule=1pt, pad at break*=1mm, colback=cellbackground, colframe=cellborder, coltitle=black, title=Box 25: Alice prepara o envio do estado que será teleportado]
\begin{Verbatim}[commandchars=\\\{\}]
\PY{n}{teleporte}\PY{o}{.}\PY{n}{cx}\PY{p}{(}\PY{n}{Alice}\PY{p}{[}\PY{l+m+mi}{0}\PY{p}{]}\PY{p}{,}\PY{n}{Alice}\PY{p}{[}\PY{l+m+mi}{1}\PY{p}{]}\PY{p}{)}
\PY{n}{teleporte}\PY{o}{.}\PY{n}{h}\PY{p}{(}\PY{n}{Alice}\PY{p}{[}\PY{l+m+mi}{0}\PY{p}{]}\PY{p}{)}
\PY{n}{teleporte}\PY{o}{.}\PY{n}{barrier}\PY{p}{(}\PY{p}{)}
\end{Verbatim}
\end{tcolorbox}

No protocolo original \cite{nielsen2002quantum,oliveira2020fisica,bennett2000quantum}, o próximo passo seria Alice realizar medidas em seus qubits e, a depender dos resultados, entrar em contato com Bob através de um canal clássico para informar as correções que Bob deve aplicar em seu estado para que o teleporte seja executado e ele consiga resgatar o estado enviado por Alice. Esse passo pode ser executado através de uma operação condicionada ao resultado das medidas de Alice. Entretanto, o IBM QE não permite a implementação desse tipo de porta condicionada a um canal clássico. Nesse caso, podemos substituí-las pela porta CNOT e Z-Controlada (construída a partir da combinação das portas Hadamard e CNOT a partir da equação (\ref{eq:15})) \cite{rabelo2018abordagem,oliveira2020fisica}. Assim, conseguimos modificar o circuito original, sem mudar seu objetivo. O Box 26 apresenta a construção da correção do protocolo de teleporte. 
\begin{tcolorbox}[breakable, size=fbox, boxrule=1pt, pad at break*=1mm, colback=cellbackground, colframe=cellborder, coltitle=black, title=Box 26: Correção do algoritmo de teleporte para o resgate do estado enviado por Alice]
\begin{Verbatim}[commandchars=\\\{\}]
\PY{n}{teleporte}\PY{o}{.}\PY{n}{h}\PY{p}{(}\PY{n}{Bob}\PY{p}{[}\PY{l+m+mi}{0}\PY{p}{]}\PY{p}{)}
\PY{n}{teleporte}\PY{o}{.}\PY{n}{cx}\PY{p}{(}\PY{n}{Alice}\PY{p}{[}\PY{l+m+mi}{0}\PY{p}{]}\PY{p}{,} \PY{n}{Bob}\PY{p}{[}\PY{l+m+mi}{0}\PY{p}{]}\PY{p}{)}
\PY{n}{teleporte}\PY{o}{.}\PY{n}{h}\PY{p}{(}\PY{n}{Bob}\PY{p}{[}\PY{l+m+mi}{0}\PY{p}{]}\PY{p}{)}
\PY{n}{teleporte}\PY{o}{.}\PY{n}{cx}\PY{p}{(}\PY{n}{Alice}\PY{p}{[}\PY{l+m+mi}{1}\PY{p}{]}\PY{p}{,} \PY{n}{Bob}\PY{p}{[}\PY{l+m+mi}{0}\PY{p}{]}\PY{p}{)}
\PY{n}{teleporte}\PY{o}{.}\PY{n}{measure}\PY{p}{(}\PY{n}{Bob}\PY{p}{,} \PY{n}{b}\PY{p}{)}
\PY{n}{teleporte}\PY{o}{.}\PY{n}{draw}\PY{p}{(}\PY{n}{output} \PY{o}{=} \PY{l+s+s1}'\PY{l+s+s1}{mpl}\PY{l+s+s1}'\PY{p}{)}
\end{Verbatim}
\end{tcolorbox}

\begin{figure}[h]
    \centering
    \includegraphics[scale=0.38]{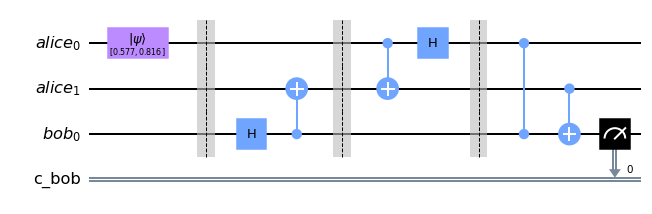}
    \caption{Representação do circuito de teleporte.} 
    \label{fig:16}
\end{figure}

A Fig. \ref{fig:16} apresenta o circuito que foi construído a partir dos Boxes 23 a 26.

\begin{figure}[h]
    \centering
    \includegraphics[scale=0.45]{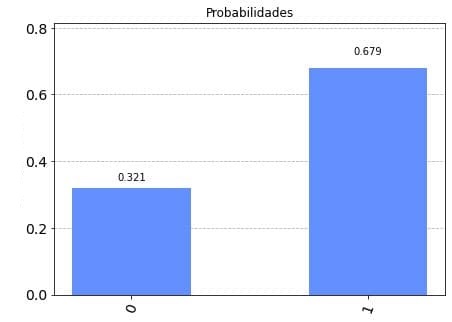}
    \caption{Distribuição de probabilidade para o algoritmo de teleporte quântico do estado $\sqrt{\frac{1}{3}}\vert{0}\rangle + \sqrt{\frac{2}{3}}\vert{1}\rangle$ executado em um processador quântico real.} 
    \label{fig:17}
\end{figure}

Ao final do processo, realizamos uma medida no qubit de Bob e obtemos a distribuição de probabilidade correspondente ao seu qubit. A fig. \ref{fig:17} apresenta a distribuição de probabilidades para o qubit de Bob após a realização do circuito da fig. \ref{fig:16} em um processador quântico real. Como pode ser visto, observa-se que o estado medido no qubit de Bob foi, em boa aproximação, o estado enviado por Alice, conforme a equação~(\ref{estadoalice}). O resultado apresentado está de acordo com a margem de erro esperada para o algoritmo de teleporte, executado no processador IBM QE de 5 qubits, conforme reportado na literatura \cite{santos2017computador,rabelo2018abordagem}.

\subsection{Algoritmo de Busca}

Um dos algoritmos mais importantes da computação quântica, e uma das principais aplicações do poder computacional do computador quântico quando comparado com um computador clássico, é o Algoritmo de Grover \cite{nielsen2002quantum,oliveira2020fisica,bennett2000quantum,castillo2020classical}.

A busca em uma lista não estruturada é um problema bastante comum nos cursos de programação. Consideremos um banco de dados não estruturado com $N$ entradas. Nosso problema é determinar o índice da entrada ($x$) do banco de dados que satisfaça algum critério de pesquisa. Para isso, definimos a função resposta ($r(x)$), uma função que mapeia classicamente as entradas do banco de dados para \texttt{True} (\texttt{0}) ou \texttt{False} (\texttt{1}), onde $r(x) = 0$ se, e somente se, $x$ satisfaz o critério de pesquisa ($x = p$), onde $p$ é o elemento procurado. Para isso, usamos uma subrotina conhecida como Oráculo, que realiza consultas à lista até encontrar o elemento $p$. Quanto mais distante o elemento procurado estiver na lista, maior o número de consultas o Oráculo precisará fazer para encontrar o elemento. Em média, a complexidade desse problema requer que o Oráculo consulte a lista $\frac{N}{2}$ vezes \cite{nielsen2002quantum,oliveira2020fisica,bennett2000quantum,castillo2020classical,figgatt2017complete}. Se o elemento estiver no final da lista, o Oráculo precisará consultá-la $N$ vezes. Logo, dizemos que o grau de complexidade desse problema é de ordem $\mathcal{O}\left(N\right)$. 
Quanticamente, o problema de busca em uma lista não estruturada é abordado no famoso Algoritmo de Grover \cite{nielsen2002quantum,oliveira2020fisica,bennett2000quantum,castillo2020classical}. Explorar a sobreposição dos estados quânticos inspecionando os $N$ itens da lista simultaneamente permite acelerar quadraticamente o problema de busca. O algoritmo de Grover é um algoritmo poderoso e sua utilidade vai além desse uso, sendo empregado como subrotina de otimização em uma grande variedade de outros algoritmos \cite{figgatt2017complete,nielsen2002quantum,magniez2007quantum,durr2006quantum,bennett1997strengths}, através do que chamamos de processo de amplificação de amplitude \cite{nielsen2002quantum}.

Como exemplo, apresentaremos a construção do Algoritmo de Grover no \textit{Qiskit} para a implementação do algoritmo de busca simples para 3 qubits \cite{nielsen2002quantum,figgatt2017complete}, em um processador quântico real. Os elementos da lista, nesse caso, são codificados na base computacional para 3 qubits $\lbrace\vert{000}\rangle , \vert{001}\rangle ,$ $\vert{010}\rangle , \vert{011}\rangle , \vert{100}\rangle , \vert{101}\rangle , \vert{110}\rangle , \vert{111}\rangle\rbrace$.  O Algoritmo de Grover é dividido em 4 partes principais: Sobreposição, Oráculo, Amplificação e a Medida \cite{nielsen2002quantum,figgatt2017complete}. 

Para inicializar os qubits em uma sobreposição igualitária, utilizamos o método apresentado na seção 3.3, aplicando a porta Hadamard em todos os qubits no processo de inicialização e obtendo o estado
\begin{eqnarray}
 \vert{\Psi_1}\rangle &=& \frac{1}{2\sqrt(2)} \left[\vert{000}\rangle + \vert{001}\rangle + \vert{010}\rangle + \vert{011}\rangle + \right. \nonumber \\ 
 && \left. + \vert{100}\rangle + \vert{101}\rangle + \vert{110}\rangle + \vert{111}\rangle\right]
 \label{grover1}.
\end{eqnarray}
Entretanto podemos implementar o código que gera o estado da equação~(\ref{grover1}) durante a inicialização do algoritmo principal. Antes, podemos construir as duas subrotinas auxiliares que formam o Algoritmo de Grover: o Oráculo e a Amplificação. 

\subsubsection{Oráculo:}

A função principal do Oráculo é marcar o elemento procurado na sobreposição \cite{nielsen2002quantum}.
Existem diferentes métodos que implementam essa subrotina \cite{nielsen2002quantum}, os dois principais são o booleano e o de inversão de fase \cite{nielsen2002quantum,figgatt2017complete}. No método booleano é necessário a presença de um qubit auxiliar (\textit{ancilla}) inicializado no estado $\vert{1}\rangle$, sendo alterado somente se a entrada para o circuito for o estado procurado. Entretanto, este método  equivalente ao método de marcação do problema de busca clássica \cite{nielsen2002quantum,figgatt2017complete} é útil para comparar o poder de computação de um computador clássico frente a um computador quântico \cite{figgatt2017complete}.

Como o objetivo desse trabalho é mostrar a aplicação de algoritmos quânticos em um processador quântico real usando o \textit{Qiskit} como uma ferramenta de ensino de computação quântica, optamos pelo método mais simples, o método de inversão de fase \cite{nielsen2002quantum,figgatt2017complete}. Nesse método não precisamos de uma \textit{ancilla}. A função do Oráculo nesse processo é identificar o elemento procurado na sobreposição equiprovável dos estados da base computacional descrita acima e adicionar uma fase negativa. Nesse contexto, o oráculo pode ser  representado pela operação unitária:
\begin{eqnarray}
U_p \vert x\rangle =
\begin{cases} 
  -\vert x\rangle & \text{se } x = p, \\
  \ \ \vert x\rangle & \text{se } x \ne p,
\end{cases} 
\label{oraculo}
\end{eqnarray}
onde $U_p$ é uma matriz diagonal que adiciona uma fase negativa à entrada que corresponde ao item procurado. $U_p$ pode ser codificado em um circuito quântico dependendo do item desejado. Os circuitos que implementam a subrotina Oráculo descrita na equação (\ref{oraculo}) em cada estado da base computacional para 3 qubits é apresentado na referência \cite{figgatt2017complete}. 

Suponhamos que o elemento procurado seja $\vert{p}\rangle =\vert{111}\rangle$. O circuito que implementa $U_p$ é a porta Z-multicontrolada que pode ser construída pela combinação da porta Toffoli e Hadamard, conforme apresentado no Box 27 \footnote{Vale destacar que não é necessário adicionar bits clássicos ao circuito pois as medidas só são executadas ao final do algoritmo principal.}.
\begin{tcolorbox}[breakable, size=fbox, boxrule=1pt, pad at break*=1mm,colback=cellbackground, colframe=cellborder,coltitle=black,title=Box 27: Iniciando o circuito e definindo o oráculo]
\begin{Verbatim}[commandchars=\\\{\}]
\PY{c+c1}{\PYZsh{} Registrando os Qubits e os Bits}

\PY{n}{q} \PY{o}{=} \PY{n}{QuantumRegister}\PY{p}{(}\PY{l+m+mi}{3}\PY{p}{,} \PY{l+s+s1}'\PY{l+s+s1}{q}\PY{l+s+s1}'\PY{p}{)}

\PY{c+c1}{\PYZsh{} Definindo a subrotina Oráculo}

\PY{n}{oraculo} \PY{o}{=} \PY{n}{QuantumCircuit}\PY{p}{(}\PY{n}{q}\PY{p}{,}\PY{n}{name} \PY{o}{=} \PY{l+s+s2}{\PYZdq{}}\PY{l+s+s2}{Oráculo}\PY{l+s+s2}{\PYZdq{}}\PY{p}{)}
\PY{n}{oraculo}\PY{o}{.}\PY{n}{h}\PY{p}{(}\PY{n}{q}\PY{p}{[}\PY{l+m+mi}{2}\PY{p}{]}\PY{p}{)}
\PY{n}{oraculo}\PY{o}{.}\PY{n}{ccx}\PY{p}{(}\PY{n}{q}\PY{p}{[}\PY{l+m+mi}{0}\PY{p}{]}\PY{p}{,}\PY{n}{}\PY{p}{[}\PY{l+m+mi}{1}\PY{p}{]}\PY{p}{,}\PY{n}{q}\PY{p}{[}\PY{l+m+mi}{2}\PY{p}{]}\PY{p}{)} 
\PY{n}{oraculo}\PY{o}{.}\PY{n}{h}\PY{p}{(}\PY{n}{q}\PY{p}{[}\PY{l+m+mi}{2}\PY{p}{]}\PY{p}{)}
\PY{n}{oraculo}\PY{o}{.}\PY{n}{draw}\PY{p}{(}\PY{n}{output} \PY{o}{=} \PY{l+s+s1}'\PY{l+s+s1}{mpl}\PY{l+s+s1}'\PY{p}{)}
\end{Verbatim}
\end{tcolorbox}

Assim, aplicando a equação~(\ref{oraculo}) na equação~(\ref{grover1}), o estado após a implementação do Box 27 será:
\begin{eqnarray}
\vert{\Psi_{oraculo}}\rangle &=& \frac{1}{2\sqrt(2)} \left[\vert{000}\rangle + \vert{001}\rangle + \vert{010}\rangle + \vert{011}\rangle + \right. \nonumber \\ 
&& \left. + \vert{100}\rangle + \vert{101}\rangle + \vert{110}\rangle - \vert{111}\rangle\right]~,
 \label{grover2}
\end{eqnarray}
adicionando uma fase negativa ao elemento $\vert{111}\rangle$.

\begin{figure}[h]
    \centering
    \includegraphics[scale=0.55]{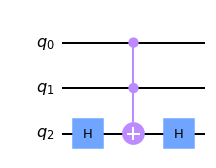}
    \caption{Oráculo para encontrar o estado $\vert{111}\rangle$}.
    \label{fig:18}
\end{figure}

A Fig. \ref{fig:18} apresenta o circuito que implementa a subrotina oráculo construído no Box 27.

Nesse ponto, mesmo tendo indicado o elemento procurado com uma fase negativa, a rotina Oráculo é insuficiente para obtermos o estado procurado, se realizarmos uma medida em nossa sobreposição, uma vez que a fase adicionada pelo Oráculo não muda a distribuição de probabilidades.

Precisamos amplificar a probablilidade do elemento procurado
$\vert{p}\rangle$ para aumentar a chance de encontrá-lo em uma medida no estado sobreposto, e reduzir as probabilidades dos demais estados da base $\vert{x}\rangle$, qualquer que seja $x \ne p$. 
Para isso, vamos usar o conhecido processo de Amplificação de Amplitude \cite{nielsen2002quantum,oliveira2020fisica,bennett2000quantum,castillo2020classical,figgatt2017complete}. 

\subsubsection{Amplificação de Amplitude}

A função da amplificação de amplitude é, como o próprio nome indica, aumentar a probabilidade do elemento marcado pelo Oráculo no estado $\vert{\Psi_{oraculo}}\rangle$, equação~(\ref{grover2}), reduzindo, consequentemente, as probabilidades dos demais itens \cite{figgatt2017complete}. Esse processo pode ser descrito em 5 subetapas \cite{nielsen2002quantum,figgatt2017complete}:
\begin{enumerate}
    \item Aplicar a porta Hadamard em todos os qubits do estado $\vert{\Psi_{oraculo}}\rangle$, equação~(\ref{grover2}), obtendo:
\begin{eqnarray}
\vert{\Psi_{1}}\rangle &=& \frac{3}{4}\vert{000}\rangle \frac{1}{4} \left[\vert{001}\rangle + \vert{010}\rangle - \vert{011}\rangle - \right. \nonumber \\ 
&& \left. - \vert{100}\rangle + \vert{101}\rangle - \vert{110}\rangle + \vert{111}\rangle\right]~;
 \label{psi1}
\end{eqnarray}

    \item Aplicar a porta X em todos os qubits do estado $\vert{\Psi_{1}}\rangle$, obtendo:
 \begin{eqnarray}
 \vert{\Psi_{2}}\rangle &=& \frac{3}{4}\vert{111}\rangle \frac{1}{4} \left[\vert{000}\rangle - \vert{001}\rangle - \vert{010}\rangle + \right. \nonumber \\ 
&&  \left. + \vert{011}\rangle - \vert{100}\rangle + \vert{101}\rangle + \vert{110}\rangle\right]~;
 \label{psi2}
\end{eqnarray}  
     \item Aplicar a porta Z-multicontrolada no estado $\vert{\Psi_{2}}\rangle$, obtendo:
     \begin{eqnarray}
 \vert{\Psi_{3}}\rangle &=& -\frac{3}{4}\vert{111}\rangle \frac{1}{4} \left[\vert{000}\rangle - \vert{001}\rangle - \vert{010}\rangle + \right. \nonumber \\
  && \left.  + \vert{011}\rangle - \vert{100}\rangle + \vert{101}\rangle + \vert{110}\rangle\right]~;
 \label{psi3}
\end{eqnarray} 
\item Aplicar novamente a porta X em todos os qubits do estado $\vert{\Psi_{3}}\rangle$, obtendo:
 \begin{eqnarray}
\vert{\Psi_{4}}\rangle &=& -\frac{3}{4}\vert{000}\rangle \frac{1}{4} \left[\vert{001}\rangle + \vert{010}\rangle - \vert{011}\rangle + \right. \nonumber \\ 
&& \left.  + \vert{100}\rangle - \vert{101}\rangle - \vert{110}\rangle + \vert{111}\rangle\right]~;
 \label{psi4}
\end{eqnarray}
\item Finalizando o processo aplicando novamente a porta Hadamard em todos os qubits do estado $\vert{\Psi_{4}}\rangle$ e obtendo o estado final
 \begin{eqnarray}
\vert{\Psi_{5}}\rangle &=& \frac{5}{4\sqrt{2}}\vert{111}\rangle +\frac{1}{4\sqrt{2}} \left[\vert{000}\rangle + \vert{001}\rangle + \vert{010}\rangle + \right. \nonumber \\ 
&& \left.  + \vert{011}\rangle + \vert{100}\rangle + \vert{101}\rangle + \vert{110}\rangle\right]~.
 \label{groverfinal}
\end{eqnarray} 
\end{enumerate}

O Box 28 apresenta a construção do circuito de amplificação, conforme descrito nessas 5 etapas:
\begin{tcolorbox}[breakable, size=fbox, boxrule=1pt, pad at break*=1mm,colback=cellbackground, colframe=cellborder, coltitle=black, title=Box 28: Criando a rotina de reflexão]
\begin{Verbatim}[commandchars=\\\{\}]
\PY{c+c1}{\PYZsh{} Definindo a subrotina Amplificação}

\PY{n}{ampl} \PY{o}{=} \PY{n}{QuantumCircuit}\PY{p}{(}\PY{n}{q}\PY{p}{,}\PY{n}{name} \PY{o}{=} \PY{l+s+s2}{\PYZdq{}}\PY{l+s+s2}{Amplificação}\PY{l+s+s2}{\PYZdq{}}\PY{p}{)}

\PY{c+c1}{\PYZsh{} Aplicar transformação | s\PYZgt{} \PYZhy{}\PYZgt{} | 00..0\PYZgt{} (porta H em todos os qbits)}
\PY{n}{ampl}\PY{o}{.}\PY{n}{h}\PY{p}{(}\PY{p}{[}\PY{n}{q}\PY{p}{[}\PY{l+m+mi}{0}\PY{p}{]}\PY{p}{,}\PY{n}{q}\PY{p}{[}\PY{l+m+mi}{1}\PY{p}{]}\PY{p}{,}\PY{n}{q}\PY{p}{[}\PY{l+m+mi}{2}\PY{p}{]}\PY{p}{]}\PY{p}{)}
\PY{c+c1}{\PYZsh{} Aplicar transformação | 00..0\PYZgt{} \PYZhy{}\PYZgt{} | 11..1\PYZgt{} (portas X)}
\PY{n}{ampl}\PY{o}{.}\PY{n}{x}\PY{p}{(}\PY{p}{[}\PY{n}{q}\PY{p}{[}\PY{l+m+mi}{0}\PY{p}{]}\PY{p}{,}\PY{n}{q}\PY{p}{[}\PY{l+m+mi}{1}\PY{p}{]}\PY{p}{,}\PY{n}{q}\PY{p}{[}\PY{l+m+mi}{2}\PY{p}{]}\PY{p}{]}\PY{p}{)}
\PY{n}{ampl}\PY{o}{.}\PY{n}{barrier}\PY{p}{(}\PY{p}{)}
\PY{c+c1}{\PYZsh{} Construindo a porta CCZ}
\PY{n}{ampl}\PY{o}{.}\PY{n}{h}\PY{p}{(}\PY{n}{q}\PY{p}{[}\PY{l+m+mi}{2}\PY{p}{]}\PY{p}{)}
\PY{n}{ampl}\PY{o}{.}\PY{n}{ccx}\PY{p}{(}\PY{n}{q}\PY{p}{[}\PY{l+m+mi}{0}\PY{p}{]}\PY{p}{,}\PY{n}{q}\PY{p}{[}\PY{l+m+mi}{1}\PY{p}{]}\PY{p}{,}\PY{n}{q}\PY{p}{[}\PY{l+m+mi}{2}\PY{p}{]}\PY{p}{)}
\PY{n}{ampl}\PY{o}{.}\PY{n}{h}\PY{p}{(}\PY{n}{q}\PY{p}{[}\PY{l+m+mi}{2}\PY{p}{]}\PY{p}{)}
\PY{n}{ampl}\PY{o}{.}\PY{n}{barrier}\PY{p}{(}\PY{p}{)}
\PY{c+c1}{\PYZsh{} Transformando o o estado de volta}

\PY{c+c1}{\PYZsh{} Aplicar transformação | 11..1\PYZgt{} \PYZhy{}\PYZgt{} | 00..0\PYZgt{} (portas X)}
\PY{n}{ampl}\PY{o}{.}\PY{n}{x}\PY{p}{(}\PY{p}{[}\PY{n}{q}\PY{p}{[}\PY{l+m+mi}{0}\PY{p}{]}\PY{p}{,}\PY{n}{q}\PY{p}{[}\PY{l+m+mi}{1}\PY{p}{]}\PY{p}{,}\PY{n}{q}\PY{p}{[}\PY{l+m+mi}{2}\PY{p}{]}\PY{p}{]}\PY{p}{)}
\PY{c+c1}{\PYZsh{} Aplicar transformação | 00..0\PYZgt{} \PYZhy{}\PYZgt{} | s\PYZgt{}  (porta H em todos os qbits)}
\PY{n}{ampl}\PY{o}{.}\PY{n}{h}\PY{p}{(}\PY{p}{[}\PY{n}{q}\PY{p}{[}\PY{l+m+mi}{0}\PY{p}{]}\PY{p}{,}\PY{n}{q}\PY{p}{[}\PY{l+m+mi}{1}\PY{p}{]}\PY{p}{,}\PY{n}{q}\PY{p}{[}\PY{l+m+mi}{2}\PY{p}{]}\PY{p}{]}\PY{p}{)}
\PY{n}{ampl}\PY{o}{.}\PY{n}{draw}\PY{p}{(}\PY{n}{output} \PY{o}{=} \PY{l+s+s1}'\PY{l+s+s1}{mpl}\PY{l+s+s1}'\PY{p}{)}
\end{Verbatim}
\end{tcolorbox}

\begin{figure}[h]
    \centering
    \includegraphics[scale=0.45]{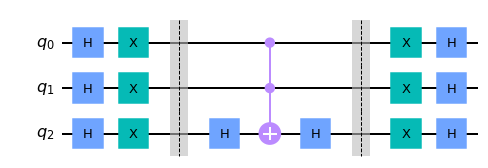}
    \caption{Circuito de amplificação de probabilidades para o algoritmo de Grover de 3 qubits.} 
    \label{fig:19}
\end{figure}

A Fig. \ref{fig:19} apresenta o circuito que implementa a subrotina de amplificação de amplitude construída no Box 28.

Assim, chegamos ao estado final da subrotina de amplificação de amplitude, equação~(\ref{groverfinal}). O algoritmo de Grover é  finalizado realizando uma medida sobre esse estado. Como pode ser visto na equação~(\ref{groverfinal}), a probailidade de encontrarmos o estado procurado $\vert{111}\rangle$ aumenta, em detrimento das probabilidades dos demais estados da base computacional para 3 qubits, caracterizando o processo de amplificação de amplitude. Se realizarmos uma medida sobre o estado $\vert{\Psi_5}\rangle$, equação~(\ref{groverfinal}), a chance de obtermos o estado  $\vert{111}\rangle$ é de aproximadamente $78,1\%$. Se quisermos aumentar ainda mais essa probabilidade, repetimos as subrotinas do oráculo e de amplificação até atingir $100\%$. De maneira geral, para uma lista não estruturada de $N$ itens, a maximização da probabilidade de encontrar o estado procurado é obtida repetindo essas duas subrotinals $\mathcal{O}\left(\sqrt{N}\right)$ vezes \cite{nielsen2002quantum,figgatt2017complete}. Por outro lado, o algoritmo clássico de busca em uma lista não estruturada precisa realizar uma média de $\frac{N}{2}$ consultas à lista para obter o elemento procurado \cite{nielsen2002quantum,figgatt2017complete}.

\subsubsection{Executando o Algoritmo:}

\begin{widetext}
 
\begin{figure}[H]
    \centering
    \includegraphics[scale=0.215]{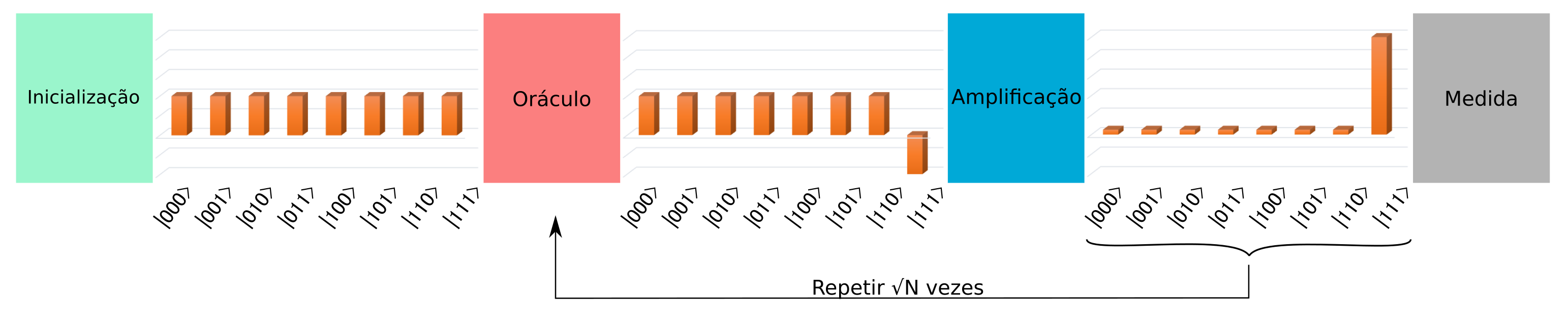}
    \caption{Esquematização de cada etapa do algoritmo de grover, mostrando a evolução das amplitudes para cada estado da base computacional para $3$ qubits.} 
    \label{fig:esquema}
\end{figure}
 
\end{widetext}

A fig.~\ref{fig:esquema} mostra uma representação esquemática para a evolução das amplitudes para cada estado da base computacional para $3$ qubits em cada etapa do algoritmo de Grover: (\textit{i}) a inicialização cria uma sobreposição igualitária de todos os estados de entrada possíveis $\lbrace\vert{000}\rangle , \vert{001}\rangle ,$ $\vert{010}\rangle , \vert{011}\rangle , \vert{100}\rangle , \vert{101}\rangle , \vert{110}\rangle , \vert{111}\rangle\rbrace$; (\textit{ii})  O oráculo marca o estado desejado de modo que a amplitude do estado procurado $\vert p\rangle$ será negativa enquanto as demais amplitudes $\vert x\rangle$ são mantidas inalteradas; (\textit{iii}) a amplificação aumenta a probabilidade de encontrarmos o estado marcado pelo oráculo; (\textit{iv}) O processo pode agora ser finalizado realizando medidas sobre todos os qubits obtendo-se o estado procurado após repetir os passos \textit{ii} e \textit{iii} $\mathcal{O}\left(\sqrt{N}\right)$ vezes. 

Vale destacar que aumentar o número de repetições dos estágios de oráculo e amplificação maximizará a amplitude da resposta correta \cite{nielsen2002quantum,oliveira2020fisica,bennett2000quantum,castillo2020classical,figgatt2017complete}. Além disso, esse algoritmo também pode ser generalizado para marcar e amplificar a amplitude de mais de um estado \cite{nielsen2002quantum,figgatt2017complete}.

\begin{figure}[h]
    \centering
    \subfigure[]{\includegraphics[scale=0.37]{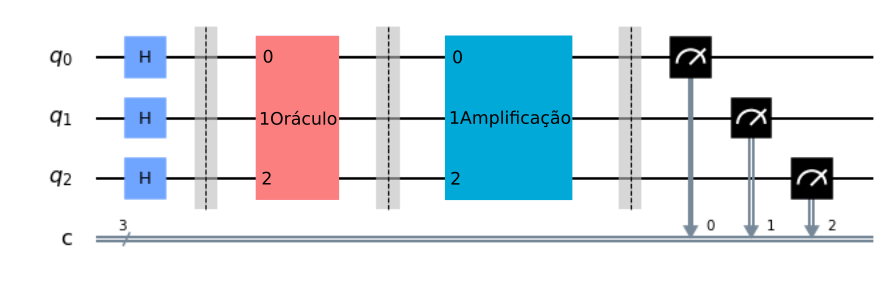}}
    \subfigure[]{\includegraphics[scale=0.26]{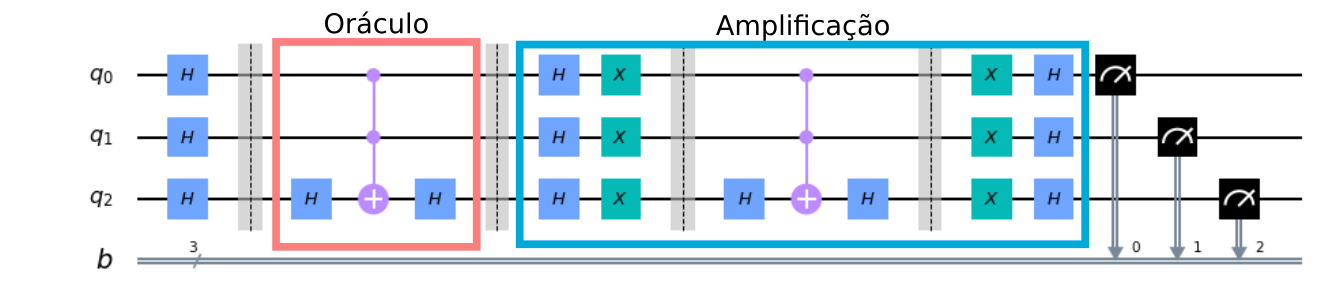}}
    \caption{Probabilidade encontrada para o algoritmo de grover.} 
    \label{fig:21}
\end{figure}

Vamos agora criar o circuito principal que implementa o algoritmo de busca através da união das subrotidas oráculo e amplificação. 
Primeiramente precisamos inicializar os qubits em uma sobreposição igualitária, como vimos na seção 3.3, aplicando a porta Hadamard em todos os qubits no processo de inicialização para criar o estado $\vert{\Psi_i}\rangle$, equação~(\ref{grover1}). Em seguida, usando o comando \texttt{grover.append()} adicionar as subrotinas Oráculo e Amplificação, criadas nos boxes 27 e 28. Finalmente realizamos as medidas e finalizamos o Algoritmo de Grover conforme apresentado no Box 29, a seguir.  
\begin{tcolorbox}[breakable, size=fbox, boxrule=1pt, pad at break*=1mm,colback=cellbackground, colframe=cellborder, coltitle=black, title=Box 29: Criando o circuito de busca]
\begin{Verbatim}[commandchars=\\\{\}]
\PY{n}{grover} \PY{o}{=} \PY{n}{QuantumCircuit}\PY{p}{(}\PY{n}{q}\PY{p}{,}\PY{n}{b}\PY{p}{)}
\PY{n}{grover}\PY{o}{.}\PY{n}{h}\PY{p}{(}\PY{p}{[}\PY{n}{q}\PY{p}{[}\PY{l+m+mi}{0}\PY{p}{]}\PY{p}{,}\PY{n}{q}\PY{p}{[}\PY{l+m+mi}{1}\PY{p}{]}\PY{p}{,}\PY{n}{q}\PY{p}{[}\PY{l+m+mi}{2}\PY{p}{]}\PY{p}{]}\PY{p}{)}
\PY{n}{grover}\PY{o}{.}\PY{n}{barrier}\PY{p}{(}\PY{p}{)}
\PY{n}{grover}\PY{o}{.}\PY{n}{append}\PY{p}{(}\PY{n}{oraculo}\PY{p}{,}\PY{n}{q}\PY{p}{)}
\PY{n}{grover}\PY{o}{.}\PY{n}{barrier}\PY{p}{(}\PY{p}{)}
\PY{n}{grover}\PY{o}{.}\PY{n}{append}\PY{p}{(}\PY{n}{ampl}\PY{p}{,}\PY{n}{q}\PY{p}{)}
\PY{n}{grover}\PY{o}{.}\PY{n}{barrier}\PY{p}{(}\PY{p}{)}
\PY{n}{grover}\PY{o}{.}\PY{n}{measure}\PY{p}{(}\PY{n}{q}\PY{p}{,}\PY{n}{b}\PY{p}{)}
\PY{n}{grover}\PY{o}{.}\PY{n}{draw}\PY{p}{(}\PY{n}{output} \PY{o}{=} \PY{l+s+s1}'\PY{l+s+s1}{mpl}\PY{l+s+s1}'\PY{p}{)}
\end{Verbatim}
\end{tcolorbox}
A fig.~ \ref{fig:21} apresenta o Algoritmo completo de Grover com as subrotinas Oráculo e Amplificação.

Finalmente, após as medidas, executamos o algoritmo de Grover em um processador quântico real e obtemos a distribuição de probabilidade correspondente. 
\begin{figure}[h]
    \centering
    \includegraphics[scale=0.5]{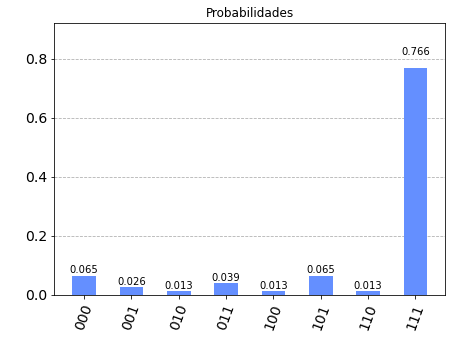}
    \caption{Distribuição de probabilidade obtida para o algoritmo de grover.} 
    \label{fig:22}
\end{figure}
Com pode ser visto, obtemos o item procurado em $76,6\%$ das $1024$ repetições. Isso significa que em uma única busca teríamos aproximadamente $76,6\%$ de chances de encontrar o elemento procurado com sucesso. Em contra partida, classicamente a chance de encontrar um item em uma lista não estruturada com $N=8$ elementos, executando somente uma consulta a lista, é de $12.5\%$, o que mostra a vantagem de usarmos propriedades quânticas como a sobreposição para o processamento da informação. Enquanto classicamente o Oráculo precisa em média realizar $N/2$ consultas a lista, quanticamente podemos encontrar o item marcado em $\sqrt{N}$ tentativas, com o método de amplificação de amplitude de Grover para o problema de busca \cite{nielsen2002quantum,figgatt2017complete}. Portanto, a junção das subrotinas Oráculo e Amplificação, para a construção do algoritmo de Grover, representam uma aceleração quadrática do problema de busca, mostrando que computadores quânticos possuem uma vantagem significativa se comparados a computadores clássicos.

\section{Conclusão}

Nesse trabalho apresentamos o kit de desenvolvimento de \textit{software} para informação quântica da IBM (\textit{Qiskit}) como uma ferramenta de trabalho para o ensino de computação e informação quântica
para os cursos de graduação em Física e áreas afins. O trabalho está estruturado na forma de um roteiro básico de sala de aula para a introdução de conceitos fundamentais da computação quântica, como qubits, portas quânticas, emaranhamento e algoritmos quânticos. Destacamos as principais condições para a construção dos programas e a sua execução em processadores quanticos reais, mostrando como essa pode ser uma ferramenta poderosa para o ensino de computação quântica de maneira prática, permitindo que os estudantes se tornem agentes ativos na construção do conhecimento. Nossos resultados estão de acordo com as previsões teóricas da literatura para os exemplos abordados, e demonstram que o \textit{Qiskit} é uma ferramenta eficaz tanto para a implementação e a análise de algoritmos quânticos simples, quanto para o desenvolvimento de softwares quânticos, atuando como uma linguagem de programação quântica  de alto nível acessível aos estudantes. 
\section*{Agradecimentos}

Os autores gostariam de agradecer a toda a equipe do \textit{IBM Research} e do \textit{Quantum Education \& Open Science at IBM Quantum} pelo acesso ao \textit{Qiskit}, e toda a comunidade do \textit{Qiskit} pelo suporte prestado ao longo do desenvolvimento desse trabalho. C. Cruz agradece a W. S. Santana pela leitura do material, a E.H.M. Maschio pelas discussões proveitosas e aos demais estudantes da disciplina  CET0448 - Tópicos Especiais III: Computação Quântica Aplicada, que mesmo não participando ativamente desse trabalho contribuiram para sua concepção.
%

\bibliographystyle{unsrt}

\begin{thebibliography}{10}

\bibitem{ibm}
IBM Quantum Experience \url{https://quantum-computing.ibm.com}.
\newblock [Acessado em: 25-Agosto-2020].

\bibitem{alves2020simulating}
{\'E}merson~M Alves, Francisco~DS Gomes, H{\'e}rcules~S Santana, and Alan~C
  Santos.
\newblock Simulating single-spin dynamics on an ibm five-qubit chip.
\newblock {\em Revista Brasileira de Ensino de F{\'\i}sica}, 42, 2020.

\bibitem{santos2017computador}
Alan~C Santos.
\newblock O computador qu{\^a}ntico da ibm e o ibm quantum experience.
\newblock {\em Revista Brasileira de Ensino de F{\'\i}sica}, 39(1), 2017.

\bibitem{rabelo2018abordagem}
Wilson~RM Rabelo and Maria L{\'u}cia~M Costa.
\newblock Uma abordagem pedag{\'o}gica no ensino da computa{\c{c}}{\~a}o
  qu{\^a}ntica com um processador qu{\^a}ntico de 5-qbits.
\newblock {\em Revista Brasileira de Ensino de F{\'\i}sica}, 40(4), 2018.

\bibitem{nielsen2002quantum}
Michael~A Nielsen and Isaac Chuang.
\newblock Quantum computation and quantum information, 2002.

\bibitem{scarani1998quantum}
Valerio Scarani.
\newblock Quantum computing.
\newblock {\em American Journal of Physics}, 66(11):956--960, 1998.

\bibitem{vedral1998basics}
Vlatko Vedral and Martin~B Plenio.
\newblock Basics of quantum computation.
\newblock {\em Progress in Quantum Electronics}, 22(1):1--39, 1998.

\bibitem{steane1998quantum}
Andrew Steane.
\newblock Quantum computing.
\newblock {\em Reports on Progress in Physics}, 61(2):117, 1998.

\bibitem{bennett2000quantum}
Charles~H Bennett and David~P DiVincenzo.
\newblock Quantum information and computation.
\newblock {\em nature}, 404(6775):247--255, 2000.

\bibitem{oliveira2020fisica}
Ivan~S Oliveira.
\newblock {\em F{\'\i}sica Qu{\^a}ntica: fundamentos formalismos e
  aplica{\c{c}}{\~o}es}, volume~1.
\newblock Editora Livraria da F{\'\i}sica, 2020.

\bibitem{terhal2018quantum}
Barbara~M Terhal.
\newblock Quantum supremacy, here we come.
\newblock {\em Nature Physics}, 14(6):530--531, 2018.

\bibitem{harrow2017quantum}
Aram~W Harrow and Ashley Montanaro.
\newblock Quantum computational supremacy.
\newblock {\em Nature}, 549(7671):203--209, 2017.

\bibitem{benenti2008quantum}
Giuliano Benenti and Giuliano Strini.
\newblock Quantum simulation of the single-particle schr{\"o}dinger equation.
\newblock {\em American Journal of Physics}, 76(7):657--662, 2008.

\bibitem{jose2013introduccao}
Marcelo~Archanjo Jos{\'e}, Jos{\'e} Roberto~Castilho Piqueira, and Roseli
  de~Deus Lopes.
\newblock Introdu{\c{c}}{\~a}o {\`a} programa{\c{c}}{\~a}o qu{\^a}ntica.
\newblock {\em Revista Brasileira de Ensino de F{\'\i}sica}, 35(1):1--9, 2013.

\bibitem{candela2015undergraduate}
D~Candela.
\newblock Undergraduate computational physics projects on quantum computing.
\newblock {\em American Journal of Physics}, 83(8):688--702, 2015.

\bibitem{fedortchenko2016quantum}
Serguei Fedortchenko.
\newblock A quantum teleportation experiment for undergraduate students.
\newblock {\em arXiv preprint arXiv:1607.02398}, 2016.

\bibitem{castillo2020classical}
Jairo~Ernesto Castillo, Yesenia Sierra, and Nelson~L Cubillos.
\newblock Classical simulation of grovers quantum algorithm.
\newblock {\em Revista Brasileira de Ensino de F{\'\i}sica}, 42, 2020.

\bibitem{perry2019quantum}
Anastasia Perry, Ranbel Sun, Ciaran Hughes, Joshua Isaacson, and Jessica
  Turner.
\newblock Quantum computing as a high school module.
\newblock {\em arXiv preprint arXiv:1905.00282}, 2019.

\bibitem{teixeira}
A.~C. Teixeira and E.~J.~R. Brandão.
\newblock Internet e democratização do conhecimento: repensando o processo de
  exclusão social.
\newblock {\em Revista Novas Tecnologias na Educação}, 1(1):1, 2003.

\bibitem{faria}
E.~V. Faria.
\newblock A tecnologia da informação e da comunicação como ferramenta para
  a construção e democratização do conhecimento.
\newblock {\em Revista Scientia FAER}, 1(1):18, 2009.

\bibitem{tappert2019experience}
Charles~C Tappert, Ronald~I Frank, Istvan Barabasi, Avery~M Leider, Daniel
  Evans, and Lewis Westfall.
\newblock Experience teaching quantum computing.
\newblock In {\em 2019 ASCUE Proceedings}. Association Supporting Computer
  Users in Education, 2019.

\bibitem{qiskit-textbook}
\url{https://qiskit.org/textbook}.
\newblock [Acessado em: 28-Agosto-2020].

\bibitem{qiskit-documentation}
\url{https://qiskit.org/documentation}.
\newblock [Acessado em: 29-Agosto-2020].

\bibitem{github-qiskit}
\url{https://github.com/Qiskit}.
\newblock [Acessado em: 28-Agosto-2020].

\bibitem{qiskit}
H{\'e}ctor Abraham, AduOffei, Rochisha Agarwal, Ismail~Yunus Akhalwaya, et~al.
\newblock Qiskit: An open-source framework for quantum computing, 2019.
\newblock \url{https://doi.org/10.5281/zenodo.2562110}.

\bibitem{larose2019overview}
Ryan LaRose.
\newblock Overview and comparison of gate level quantum software platforms.
\newblock {\em Quantum}, 3:130, 2019.

\bibitem{qiskit-tutorials}
\url{https://github.com/Qiskit/qiskit-tutorials}.
\newblock [Acessado em: 15-Janeiro-2021].

\bibitem{python}
Guido Van~Rossum and Fred~L. Drake.
\newblock {\em Python 3 Reference Manual}.
\newblock CreateSpace, Scotts Valley, CA, 2009.

\bibitem{oliphant2007python}
Travis~E Oliphant.
\newblock Python for scientific computing.
\newblock {\em Computing in Science \& Engineering}, 9(3):10--20, 2007.

\bibitem{van2011python}
Guido Van~Rossum and Fred~L Drake.
\newblock {\em The python language reference manual}.
\newblock Network Theory Ltd., 2011.

\bibitem{kadiyala2017applications}
Akhil Kadiyala and Ashok Kumar.
\newblock Applications of python to evaluate environmental data science
  problems.
\newblock {\em Environmental Progress \& Sustainable Energy}, 36(6):1580--1586,
  2017.

\bibitem{harris2020array}
Charles~R Harris, K~Jarrod Millman, St{\'e}fan~J van~der Walt, Ralf Gommers,
  Pauli Virtanen, David Cournapeau, Eric Wieser, Julian Taylor, Sebastian Berg,
  Nathaniel~J Smith, et~al.
\newblock Array programming with numpy.
\newblock {\em Nature}, 585(7825):357--362, 2020.

\bibitem{kluyver2016jupyter}
Thomas Kluyver, Benjamin Ragan-Kelley, Fernando P{\'e}rez, Brian~E Granger,
  Matthias Bussonnier, Jonathan Frederic, Kyle Kelley, Jessica~B Hamrick, Jason
  Grout, Sylvain Corlay, et~al.
\newblock Jupyter notebooks-a publishing format for reproducible computational
  workflows.
\newblock In {\em ELPUB}, pages 87--90, 2016.

\bibitem{jupyter}
\url{https://jupyter.org}.
\newblock [Acessado em: 20-Agosto-2020].

\bibitem{glick2018using}
Ben Glick and Jens Mache.
\newblock Using jupyter notebooks to learn high-performance computing.
\newblock {\em Journal of Computing Sciences in Colleges}, 34(1):180--188,
  2018.

\bibitem{cardoso2018using}
Alberto Cardoso, Joaquim Leit{\~a}o, and C{\'e}sar Teixeira.
\newblock Using the jupyter notebook as a tool to support the teaching and
  learning processes in engineering courses.
\newblock In {\em International Conference on Interactive Collaborative
  Learning}, pages 227--236. Springer, 2018.

\bibitem{zuniga2020digital}
Arturo Z{\'u}{\~n}iga-L{\'o}pez and Carlos Avil{\'e}s-Cruz.
\newblock Digital signal processing course on jupyter--python notebook for
  electronics undergraduates.
\newblock {\em Computer Applications in Engineering Education},
  28(5):1045--1057, 2020.

\bibitem{perkel2018jupyter}
Jeffrey~M Perkel.
\newblock Why jupyter is data scientists' computational notebook of choice.
\newblock {\em Nature}, 563(7732):145--147, 2018.

\bibitem{hussain2018introducing}
Zahid Hussain and Muhammad~Siyab Khan.
\newblock Introducing python programming for engineering scholars.
\newblock {\em INTERNATIONAL JOURNAL OF COMPUTER SCIENCE AND NETWORK SECURITY},
  18(12):26--33, 2018.

\bibitem{anaconda}
\textit{Anaconda Software Distribution}. Computer software. Vers. 2-2.4.0.
  Anaconda, Nov. 2016. Web:\url{https://anaconda.com}.
\newblock [Acessado em: 20-Agosto-2020].

\bibitem{qiskit-swift}
\url{https://github.com/qiskit-community/qiskit-swift}.
\newblock [Acessado em: 15-Janeiro-2021].

\bibitem{qiskit-js}
\url{https://github.com/qiskit-community/qiskit-js}.
\newblock [Acessado em: 15-Janeiro-2021].

\bibitem{hunter2007matplotlib}
John~D Hunter.
\newblock Matplotlib: A 2d graphics environment.
\newblock {\em Computing in science \& engineering}, 9(3):90--95, 2007.

\bibitem{vedral2006introduction}
Vlatko Vedral et~al.
\newblock {\em Introduction to quantum information science}.
\newblock Oxford University Press on Demand, 2006.

\bibitem{gyongyosi2019survey}
Laszlo Gyongyosi and Sandor Imre.
\newblock A survey on quantum computing technology.
\newblock {\em Computer Science Review}, 31:51--71, 2019.

\bibitem{pedroni2008digital}
Volnei~A Pedroni.
\newblock {\em Digital electronics and design with VHDL}.
\newblock Morgan Kaufmann, 2008.

\bibitem{griffiths2018introduction}
DJ~Griffiths.
\newblock {\em Mec{\^a}nica Qu{\^a}ntica, 2{\textordfeminine}
  Edi{\c{c}}{\~a}o}.
\newblock Editora Pearson Education, 2011.

\bibitem{whitesitt2012boolean}
J~Eldon Whitesitt.
\newblock {\em Boolean algebra and its applications}.
\newblock Courier Corporation, 2012.

\bibitem{horodecki}
Ryszard Horodecki, Pawe{\l} Horodecki, Micha{\l} Horodecki, and Karol
  Horodecki.
\newblock Quantum entanglement.
\newblock {\em Reviews of modern physics}, 81(2):865, 2009.

\bibitem{figgatt2017complete}
Caroline Figgatt, Dmitri Maslov, KA~Landsman, Norbert~Matthias Linke, Shantanu
  Debnath, and C~Monroe.
\newblock Complete 3-qubit grover search on a programmable quantum computer.
\newblock {\em Nature communications}, 8(1):1--9, 2017.

\bibitem{magniez2007quantum}
Fr{\'e}d{\'e}ric Magniez, Miklos Santha, and Mario Szegedy.
\newblock Quantum algorithms for the triangle problem.
\newblock {\em SIAM Journal on Computing}, 37(2):413--424, 2007.

\bibitem{durr2006quantum}
Christoph D{\"u}rr, Mark Heiligman, Peter HOyer, and Mehdi Mhalla.
\newblock Quantum query complexity of some graph problems.
\newblock {\em SIAM Journal on Computing}, 35(6):1310--1328, 2006.

\bibitem{bennett1997strengths}
Charles~H Bennett, Ethan Bernstein, Gilles Brassard, and Umesh Vazirani.
\newblock Strengths and weaknesses of quantum computing.
\newblock {\em SIAM journal on Computing}, 26(5):1510--1523, 1997.

\end{thebibliography}

\end{document}